%% file: main.tex
\documentclass[%
 reprint, showkeys,
superscriptaddress,
 amsmath,amssymb,
 aps,
pra
]{revtex4-2}

\usepackage{graphicx}
\usepackage{dcolumn}
\usepackage{bm}
\usepackage{hyperref}
\usepackage{breakurl}
\usepackage{subcaption}
\usepackage[justification=centerlast]{caption} 
\usepackage{xcolor} 
\usepackage{fullpage} 
\usepackage[english]{babel}
\usepackage{tikz}
\usetikzlibrary{external}
\usetikzlibrary{quantikz}
\usepackage[T1]{fontenc}
\usepackage{tabularx}
\usepackage[normalem]{ulem}
\usepackage{soul}

\makeatletter
\let\ORIbbl@fixname\bbl@fixname
\def\bbl@fixname#1{%
  \@ifundefined{languagealias@\expandafter\string#1}
    {\ORIbbl@fixname#1}
    {\edef\languagename{\@nameuse{languagealias@#1}}}%
}
\newcommand{\definelanguagealias}[2]{%
  \@namedef{languagealias@#1}{#2}%
}

\DeclareMathOperator{\Tr}{Tr}
\newcommand{\ketbra}[2]{| #2 \rangle\langle #1 |}
\makeatother
\definelanguagealias{en}{english}
\definelanguagealias{en-GB}{english}

\begin{document}

\preprint{APS/123-QED}

\title{Entropy Density Benchmarking of Near-Term Quantum Circuits} 

\author{Marine Demarty}
 \email{corresponding author\\marine.demarty@gmail.com}
 \affiliation{%
 School of Informatics, University of Edinburgh,\\
 10 Crichton Street, Edinburgh EH8 9AB, United Kingdom
}%
\author{James Mills}%
 \affiliation{%
 School of Informatics, University of Edinburgh,\\
 10 Crichton Street, Edinburgh EH8 9AB, United Kingdom
}%
\affiliation{%
Quandela,\\%
7 Rue Léonard de Vinci, 91300 Massy, France} 
\author{Kenza Hammam}
\affiliation{%
 School of Informatics, University of Edinburgh,\\
 10 Crichton Street, Edinburgh EH8 9AB, United Kingdom
}%
\author{Raúl García-Patrón}
\affiliation{%
 School of Informatics, University of Edinburgh,\\
 10 Crichton Street, Edinburgh EH8 9AB, United Kingdom
}%
\affiliation{Phasecraft Ltd.,\\
77-79 Charlotte Street, London W1T 4PW, United Kingdom}


\begin{abstract}
Understanding the limitations imposed by noise on current and next-generation quantum devices is a crucial step toward demonstrating practical quantum advantage. In this work, we investigate the accumulation of entropy density as a benchmark to monitor the performance of quantum processing units (QPUs). We provide a proof-of-principle demonstration of our methodology which entails developing simple heuristic models of how entropy accumulates, testing them against real QPU data, and finally using these models to determine a circuit volume threshold above which quantum advantage is unattainable. Monitoring entropy density not only offers an alternative approach that complements existing circuit-level benchmarking techniques, but, more importantly, it bridges the gap between circuit-level and application-level benchmarking protocols. In particular, our heuristic model of entropy accumulation allows us to outperform existing techniques that bound the circuit-size threshold for quantum advantage.
\end{abstract}


\maketitle

\section{\label{sec:Introduction}Introduction}

One of the main limitations of current and near-term quantum devices remains the imperfect control of the qubits and the challenge of isolating the system from its environment, leading to noise levels beyond those required to reach fault-tolerant quantum computation \cite{gottesman1996class, gottesman_introduction_2009, roffe2019quantum}. This regime is known as noisy intermediate-scale quantum (NISQ) computation. Even if noise greatly limits the computational power of such computers, their practical applications are worth investigating as their size is already above the limit of 50 qubits---which is typically when error-free quantum devices become highly nontrivial to simulate by a classical digital computer \cite{preskill_quantum_2018, bharti2022noisy}.

To harness the full power of NISQ technology, however, one must look for tasks that are suitable for intermediate-scale devices with high error rates; this means, for instance, designing quantum algorithms that are resilient to noise, and finding tests of quantum advantage that do not require more than a few hundred qubits. Examples of such tasks include random circuit sampling \cite{arute_quantum_2019} and Gaussian boson sampling \cite{zhong_quantum_2020, zhong_phase-programmable_2021}, which have been the target of recent quantum supremacy claims. However, these sampling tasks have very limited useful application (see, e.g., \cite{yu_universal_2023, arrazola_using_2018, aaronson_certified_2023}).

A popular way to design applications for NISQ devices is to reformulate a given computational task as an optimization problem which can be solved using so-called variational quantum algorithms (VQAs) \cite{cerezo_variational_2021}. These hybrid quantum-classical algorithms are believed to be good candidates to implement on NISQ devices affected by errors and low coherence times. Solving combinatorial optimization problems~\cite{farhi_quantum_2014} or finding the ground-state energy of a Hamiltonian~\cite{peruzzo_variational_2014} are examples among an extensive literature on the topic \cite{cerezo_variational_2021}. Whether quantum advantage can be achieved in the near term on \textit{practical} tasks remains a question of scientific debate, as seen after the recent simulation of the behavior of a magnetic material on a 127-qubit quantum processor \cite{kim_evidence_2023} and its subsequent rebuttals (see, e.g., Refs. \cite{tindall_efficient_2024, begusic_fast_2024}).

To understand and evaluate the potential of noisy quantum devices to solve specific tasks, it is crucial to be able to assess the effect of errors on performance. With this in mind, a wide range of benchmarking techniques have been developed. Those fall into three main categories: gate-, circuit-, and application-level protocols. Gate-level benchmarking protocols aim at characterizing the performance of low-level components such as gates and sequences of gates (e.g., randomized benchmarking \cite{magesan_characterizing_2012}, gateset tomography \cite{nielsen_gate_2021}). Circuit-level protocols are concerned with assessing the performance of a quantum processing unit (QPU) when running specific classes of circuits (e.g., cross-entropy benchmarking \cite{arute_quantum_2019}, quantum volume \cite{cross_validating_2019}). Finally, application-level benchmarking techniques characterize the performance of a QPU when solving practical problems (e.g., Q-score for Max-Cut with quantum approximate optimization algorithm (QAOA) \cite{martiel_benchmarking_2021}, fermionic depth for many-body problems \cite{dallaire-demers_application_2020}). However, to the best of our knowledge, those benchmarking categories are relatively isolated and there are no benchmarks that exploit benchmarking tools of one level to provide information at a higher level.

We believe entropy density benchmarking provides an excellent bridge between circuit-level and application-level benchmarking protocols, enabling us to estimate the circuit size above which quantum advantage is unattainable for solving a given problem of interest. A significant part of the errors affecting quantum circuits, such as stochastic Pauli channels, leads to an increase of entropy. Intuitively, the accumulation of entropy is bad for the performance of the system, as the entropy quantifies closeness to a maximally mixed quantum state. The link between entropy accumulation and the performance of variational quantum algorithms was recently formalized in Ref. \cite{stilck_franca_limitations_2021}, providing a rigorous formalism that bridges the gap between circuit- and application-level benchmarking.
We envision the use of entropy density as a central tool of a benchmarking methodology that consists of three main steps: (i) combination of analytical and numerical methods to build simple yet effective heuristic models of entropy accumulation, (ii) validation and adjustment after comparison with real hardware entropy data, and (iii) bounding the circuit volume accessible before losing any potential quantum advantage. 

In this work, we aim to provide a proof-of-principle demonstration of this benchmarking methodology allowing to bridge the gap between circuit-level and application-level benchmarking protocols, enabling us to derive more accurate thresholds than in Ref. \cite{stilck_franca_limitations_2021} on the circuit size above which quantum advantage is unattainable.

The structure of this paper is as follows. In Sec. \ref{sec:Classical_Simulation}, we analyze numerically the Renyi-2 entropy scaling of the output register of a hardware-efficient VQA circuit with the circuit size (width and depth) and noise parameters, motivating the use of an analytical model for entropy accumulation in such circuits based on global depolarizing noise. In Sec. \ref{sec:Experimental_Study}, we test the model against entropy accumulation data collected on a real superconducting device and discuss further refinements of the model to bring it closer to actual hardware observations. We then argue that global depolarizing, despite some limitations, still provides a simple heuristic approach to decide whether quantum advantage is attainable with a typical superconducting NISQ device for a specific optimization problem in Sec. \ref{sec:App1_Quantum_Advantage_Benchmarking}. Finally, we summarize the main results of this paper and discuss open questions and possible future directions to this work in Sec. \ref{sec:Conclusion}.

\section{\label{sec:Classical_Simulation}Classical Simulation}

\begin{figure}[h]
    \centering
    \resizebox{0.48\textwidth}{!}{%
    \begin{quantikz}[thin lines]
    \lstick{$\ket{0}$} &  \gate{R_X(\theta_1)} &\gate{R_Y(\theta_6)} & \ctrl{1} & \qw \slice{end of layer 1} &  \gate{R_X(\theta_{11})} &\gate{R_Y(\theta_{16})} & \ctrl{1} & \qw \slice{end of layer 2} & \qw &\\
    \lstick{$\ket{0}$} & \gate{R_X(\theta_2)} &\gate{R_Y(\theta_7)} & \ctrl{} & \ctrl{1} &  \gate{R_X(\theta_{12})} &\gate{R_Y(\theta_{17})} & \ctrl{} & \ctrl{1} & \qw &\\
    \lstick{$\ket{0}$} & \gate{R_X(\theta_3)} &\gate{R_Y(\theta_8)} & \ctrl{1} & \ctrl{} &  \gate{R_X(\theta_{13})} &\gate{R_Y(\theta_{18})} & \ctrl{1} & \ctrl{} & \qw & \rstick{\dots}\\
    \lstick{$\ket{0}$} & \gate{R_X(\theta_4)} &\gate{R_Y(\theta_9)} & \ctrl{} & \ctrl{1} &  \gate{R_X(\theta_{14})} &\gate{R_Y(\theta_{19})} & \ctrl{} & \ctrl{1} & \qw &\\
    \lstick{$\ket{0}$} & \gate{R_X(\theta_5)} &\gate{R_Y(\theta_{10})} & \qw & \ctrl{} &  \gate{R_X(\theta_{15})} &\gate{R_Y(\theta_{20})} & \qw & \ctrl{} & \qw &
    \end{quantikz}
    }%
    \caption{First two circuit layers of a five-qubit hardware-efficient parameterized quantum circuit $U(\Vec{\theta})$. Each circuit layer is composed of a single-qubit $R_X(\theta)$ rotation followed by an $R_Y(\theta)$ rotation
    and a final layer of $CZ$ gates between nearest neighbors. The angles $\theta_i$ are drawn from the uniform distribution over $[0,2\pi)$.}
    \label{fig:ideal_hardware_efficient_param_circuit}
\end{figure}
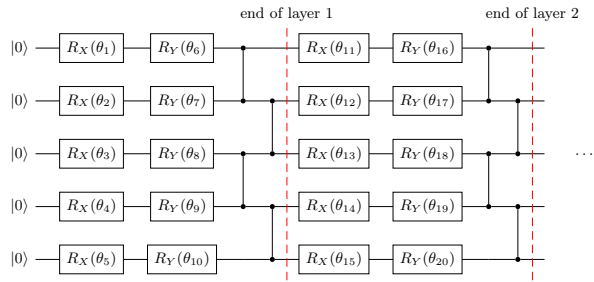

We are interested in how the entropy of the quantum register evolves in a variational quantum circuit as layers of gates are applied to the input state, causing errors to accumulate, and in the dependency of the entropy on the circuit width (number of qubits). 
As a worked example, we consider a typical hardware-efficient VQA ansatz: each layer of this circuit consists of the concatenation of random single-qubit rotations about the $X$ axis $R_X(\theta) \coloneqq \exp(-i \frac{\theta}{2} X)$, with random single-qubit rotations about the $Y$ axis $R_Y(\theta) \coloneqq \exp(-i \frac{\theta}{2} Y)$, followed by a set of nearest-neighbor $CZ$ gates distributed into two layers of nonintersecting gates, as shown in Fig.~\ref{fig:ideal_hardware_efficient_param_circuit}.

\subsection{Entropy density accumulation under local depolarizing noise}%

We start by numerically investigating the entropy accumulation of our circuit of interest for different circuit widths and number of circuit layers (depth), under a commonly used local depolarizing noise model. Let $p_1$ ($p_2$) be the depolarizing probability for single-qubit (two-qubit) gates. The noise model we choose for the numerical simulation consists of applying a single-qubit (two-qubit) depolarizing channel with depolarizing probability $p_1$ ($p_2$) after each single-qubit (two-qubit) gate of the quantum circuit from Fig.~\ref{fig:ideal_hardware_efficient_param_circuit}. For a single-qubit state $\rho$, our single-qubit depolarizing channel is $\mathcal{N}_{p_1}(\rho) \coloneqq (1 - p_1) \rho + p_1 I/2$, while for a two-qubit state $\rho$, the two-qubit depolarizing channel considered is $\mathcal{N}_{p_2}(\rho) \coloneqq (1 - p_2) \rho + p_2 I/4$.

We use the software development kit qiskit \cite{noauthor_qiskit_nodate} to
calculate the entropy of the quantum register at depth $D$ by evolving the density matrix of $n$ qubits starting with the input state $\rho_0 \coloneqq (\ket{0}\bra{0})^{\otimes n}$. 
For each circuit layer, we compute the second-order Renyi entropy defined as 
\begin{equation}
    S^{(2)}(\rho_{D}) = -\log_2(\Tr[\rho_{D}^2])\,,
\end{equation}
where $\rho_{D}$ is the output density matrix (after applying $D$ layers of gates). 
\begin{figure*}
    \centering
    \includegraphics[width=\textwidth]{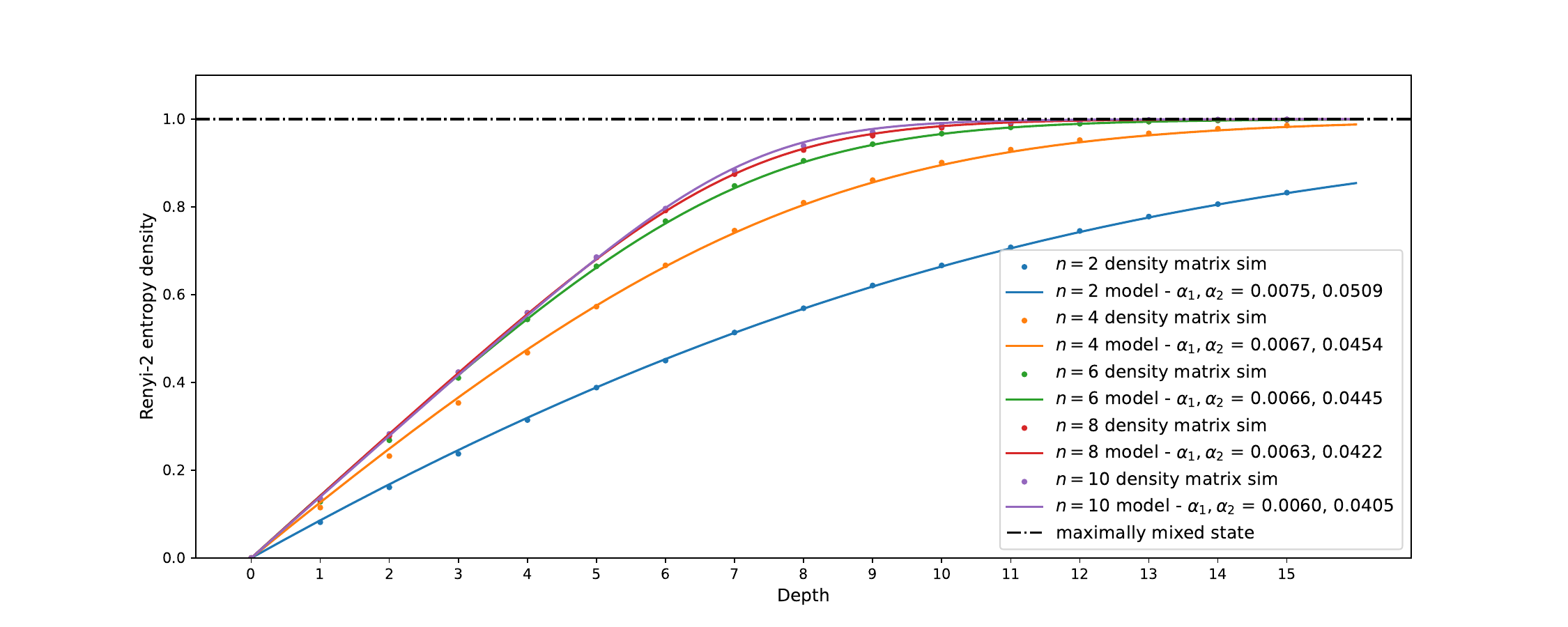}
    \caption{\textbf{Numerical simulation and heuristic model.} We simulate a single VQA circuit, represented in Fig.~\ref{fig:ideal_hardware_efficient_param_circuit}, with randomly selected parameters (random seed \texttt{np.random.seed(837)}) affected by local depolarizing noise ($p_1=0.008$ and $p_2=0.054$ from calibration data of Rigetti's QPU, Aspen-M-3). The dots characterize the Renyi-2 entropy density evolution of the quantum register after each circuit layer for different number of qubits, $n$ (see color label). The black horizontal dash-dotted line corresponds to the entropy density of the maximally mixed state $\sigma_0\coloneqq I/2^n$. 
    The solid lines correspond to interpolations using a global depolarizing heuristic model from Eq.~\eqref{eq:purity-model-from-class-sim} with fitting parameters $\alpha_1$ and $\alpha_2$ ($\alpha_i\approx p_i$), as detailed in Sec. \ref{subsubsec:analytical-model}.}
    \label{fig:numerical_densmat_simulation_with_fit}
\end{figure*}
The obtained evolution of the Renyi entropy density $S^{(2)}(\rho_{D})/n$ for different numbers of qubits, $n$ (widths), and different depths can be found in Fig.~\ref{fig:numerical_densmat_simulation_with_fit}, where for $p_1$ and $p_2$ we take deduced values from QPU calibration data (see Appendix \ref{app:calibration_data_Aspen-M-3}).
We focus our interest on entropy density in order to allow a meaningful comparison between circuits of different width with the additional benefit of potentially inferring universal rules independently of the system size. Note that for every circuit size and depth, the dotted results correspond to the classical simulation using local depolarizing noise, while the solid lines correspond to an interpolation given by a global depolarizing model discussed in the section below.

As expected, local depolarizing noise causes the system to converge toward the maximally mixed state as the error occurrence increases, approaching a density of $1$ at high depth. Interestingly, the larger the width $n$ of the circuit, the faster the entropy density converges, where for large systems ($n \rightarrow \infty$) the speed of convergence of the Renyi entropy density seems to converge to an asymptotic bound, as one can clearly see from Fig. \ref{fig:numerical_densmat_simulation_with_fit}.

\subsection{Global depolarizing heuristic model}

\subsubsection{Analytical model}\label{subsubsec:analytical-model}

We wish to build a heuristic model for the evolution of the second-order Renyi entropy as a function of the quantum circuit size, i.e., depth and width. 
Recent random circuit sampling quantum supremacy demonstrations \cite{dalzell2024random}, as well as some error mitigation techniques run on superconducting hardware \cite{urbanek2021mitigating}, have relied on a seemingly successful global depolarizing noise (or white noise) assumption. It is therefore natural to explore the approximation of the chosen local depolarizing noise model described previously by a global depolarizing noise channel applied after the last layer of the ideal quantum circuit $U_D(\Vec{\theta})$; i.e., the output of our noisy VQA circuit can be written as
\begin{equation}
    \rho_{D} = (1-P) U_D(\Vec{\theta})\rho_0 U_D(\Vec{\theta})^\dagger + P \frac{I}{2^n} \,,
\end{equation}
where $P$ is the global depolarizing probability, and $\rho_0 = (\ket{0}\bra{0})^{\otimes n}$ is the input state. The analytical model for the purity then reads
\begin{small}
\begin{equation}
    \Tr[\rho_{D}^2] = (1 - 2^{-n})((1-P)^2 - 1) + 1 \,.\label{eq:model-function-of-global-noise-param}
\end{equation}
\end{small}

To link the global depolarizing probability $P$ of the model with the circuit size $(n, D)$ and local noise parameters $(p_1, p_2)$, we exploit the standard assumption that the probability of no error occurring in the circuit under local depolarizing noise should match the probability of no error occurring in the model with global depolarizing noise. This yields
\begin{align}
    1 - P &= (1 - p_1)^{|\text{1Q gates}|} (1 - p_2)^{|\text{2Q gates}|}\,,\\
    &= e^{-\alpha_1 2nD} e^{-\alpha_2(n-1)D} \,,\label{eq:link-global-noise-param-and-fitting-params}
\end{align}
where $|\text{1Q gates}|$ ($|\text{2Q gates}|$) is the number of one-qubit (two-qubit) gates of the circuit, corresponding to $2n$ and $n-1$, respectively, for the circuit of Fig. \ref{fig:ideal_hardware_efficient_param_circuit}. 

We have introduced the parameters $\alpha_1 \coloneqq \ln((1 - p_1)^{-1})$ and $\alpha_2 \coloneqq \ln((1 - p_2)^{-1})$, which are expected to model the depolarizing probabilities $(p_1, p_2)$ in the low-noise limit: $\alpha_1 \approx p_1$ and $\alpha_2 \approx p_2$. Substituting Eq.~\eqref{eq:model-function-of-global-noise-param} with Eq.~\eqref{eq:link-global-noise-param-and-fitting-params}, we get the following model for the purity:

\begin{small}
\begin{equation}
    \Tr[\rho_{D}^2] = (1 - 2^{-n})(e^{-2(2\alpha_1 n+\alpha_2(n-1))D} - 1) + 1 \,,\label{eq:purity-model-from-class-sim}
\end{equation}
\end{small}
with fitting parameters $\alpha_1$ and $\alpha_2$. 

We subsequently performed a nonlinear least-squares fitting of the purity data obtained from numerical simulation; the corresponding Renyi-2 entropy density data (dots) with the Renyi-2 entropy density of the fitted purity model are shown in Fig.~\ref{fig:numerical_densmat_simulation_with_fit}.
We explored three different strategies for the use of the free parameters $\alpha_i$: (i) leaving both parameters free $\alpha_i$, (ii) fixing $\alpha_1=0$, and
(iii) assuming $\frac{\alpha_1}{\alpha_2} = \frac{p_1}{p_2}$, where the last two assumptions are different approaches to model the gap between single-qubit and two-qubit gate errors. If all three assumptions lead to very similar qualities of interpolation, the first leads to parameters $\alpha_1$ and $\alpha_2$ that have little connection with the physical values $p_1$ and $p_2$. Therefore, we use the constraint $\frac{\alpha_1}{\alpha_2} = \frac{p_1}{p_2}$ in all the fittings presented in this paper, to guarantee the same order of magnitude between all pairs ($\alpha_i,p_i$).

Interestingly, a global depolarizing model provides a good heuristic description of the entropy accumulation behavior at considered noise levels and for small systems without having to average over the whole family of circuits (variational gate parameters); it achieves a good fit already for a given specific circuit.

\subsubsection{Asymptotic behaviors}\label{subsubsec:asymptotic_behaviors}

It is easy to see that for the limit $D \rightarrow \infty$, the global depolarizing model leads to $\Tr[\rho_{D}^2] \sim 1/2^n$, which results in $S^{(2)}(\rho_{D})/n \sim 1$, reaching the entropy density of the maximally mixed state as observed in Fig.~\ref{fig:numerical_densmat_simulation_with_fit}.

For all system sizes, at small enough circuit depth, one can observe in Fig.~\ref{fig:numerical_densmat_simulation_with_fit}, for local depolarizing noise,
that the Renyi entropy density growth is well approximated by a linear function with the circuit depth. 
Moreover, in the limit of large system sizes, $n \rightarrow \infty$, the Renyi entropy density growth seems to converge to an upper bound.

A similar feature can be derived  for the global depolarizing model as the Renyi-2 entropy density bound
\begin{small}
\begin{equation}
    \lim_{n \rightarrow + \infty}\frac{S^{(2)}(\rho_{D})}{n} = \begin{cases}
        \frac{2(2\alpha_1 + \alpha_2)}{\ln(2)}D & \text{ if } D \leq \frac{\ln(2)}{2(2\alpha_1 + \alpha_2)} \\
        1 & \text{ if } D > \frac{\ln(2)}{2(2\alpha_1 + \alpha_2)} 
    \end{cases}\,,
\end{equation}
\end{small}
which defines the depth threshold 
\begin{equation}
    D^* = \frac{\ln(2)}{2(2\alpha_1 + \alpha_2)} \label{eq:depth_threshold_maximised_entropy_large_systems}
\end{equation}
above which the output is indistinguishable from the maximally mixed state. For instance, taking typical QPU values for noise parameters $\alpha_1 = p_1 = 10^{-4}$ and $\alpha_2 = p_2 = 10^{-3}$~\cite{acharya2024quantum}, we get a depth threshold $D^* = 288$, beyond which the quantum device's output is no better than a random guess.

It is important to note that the obtained bound to the Renyi-2 entropy density for the global depolarizing noise model in the limit of large system sizes assumes fixed $\alpha_1$ and $\alpha_2$ values. When the heuristic model is fitted on the local depolarizing simulations, we can see from Fig. \ref{fig:numerical_densmat_simulation_with_fit} that the fitting parameters $\alpha_1$ and $\alpha_2$ vary with the system size $n$. Nonetheless, assuming the fitting parameters saturate to constant values as $n\rightarrow \infty$ as Fig. \ref{fig:numerical_densmat_simulation_with_fit} suggests, this asymptotic result for global depolarizing noise captures the local depolarizing behavior.

\section{\label{sec:Experimental_Study}Analysis on a real QPU}

We are now ready to study entropy accumulation as a function of circuit size on an actual quantum processing unit, and explore whether our model for the Renyi-2 entropy from Eq.~\eqref{eq:purity-model-from-class-sim} matches QPU results.
In Sec. \ref{subsubsec:experimental_study_of_entropy_accumulation}, we provide detailed analytical, numerical, and QPU data over a superconducting platform using classical shadows, a technique that has proven extremely useful for the estimation of expectation values on NISQ devices. 
Despite not being scalable, the classical shadows protocol provides a good compromise for small circuit sizes like those explored in this work and more generally for systems made of a handful of qubits. We observe that the QPU data diverge from a simple global depolarizing model and are able to explain most of this gap by incorporating a model of $T_1$
relaxation at the gate and measurement level in Sec. 
\ref{Sec Exp - Discussion results}.

\subsection{Classical Shadows}

\subsubsection{Background}

In their 2020 paper, Huang \textit{et al.} introduce a new method for efficient estimation of many properties of a quantum state $\rho$ known as the classical shadows protocol \cite{huang_predicting_2020}. Instead of performing full tomography of the quantum state, which requires a number of measurements that scale exponentially with the system size $n$, they proposed to obtain several classical snapshots of the quantum state by measuring the state in $M$ randomly chosen bases $K$ times. Classical postprocessing of the measurement outcomes then allows to deduce the expectation value of $N_{obs}$ observables of interest $\Tr[O_i\rho]$, in a sample efficient manner for specific families of observables.

Huang \textit{et al.} propose two versions of this protocol: (i) random single-qubit Pauli measurements and (ii) random $n$-qubit Clifford circuit measurements.
Pauli measurements are particularly efficient for predicting a large collection of local observables, and only require short-depth measurement circuits, which are very suitable for NISQ devices. On the other hand, $n$-qubit Clifford measurements should be preferred for efficiently predicting many global observables with a bounded Hilbert-Schmidt norm; the main drawback is that gates of order $n^2/\log(n)$ are needed for their implementation, resulting in circuits that are less practical for current quantum devices.

More precisely, to estimate $N_{obs}$ observables $O_i$ with additive error $\epsilon$, the number $M$ of randomly chosen bases, measured $K$ times, needs to scale as
\begin{equation}
    O(\log(N_{obs})\max_i ||O_i||_{shadow}^2/{\epsilon}^2),
\end{equation} 
where $||O_i||_{shadow}$ is the so-called shadow norm, which is upper bounded by $\max_i 4^{k_i}||O_i||_{\infty}^2$ in scenario (i) of single-qubit Clifford measurements, and by $\max_i \Tr[O_i^2]$ in scenario (ii) of global Clifford measurements (see Ref. \cite{huang_predicting_2020}), where $k_i$ is the locality of observable $O_i$ and $||.||_\infty$ is the operator norm.

\subsubsection{Renyi-2 entropy estimation}

The above framework also extends to nonlinear functions of $\rho$, such as quadratic terms $\Tr[O_i\rho \otimes \rho]$. By noting that the second-order Renyi entropy of a quantum state, $S^{(2)}(\rho)$, is a function of the purity $\Tr[\rho^2]$, which can be rewritten as
\begin{equation}
    \Tr[\rho^2] = \Tr[S \rho \otimes \rho] \,,\label{eq:purity-swap-gate-relation}
\end{equation}
where $S$ is the SWAP matrix, one can use the classical shadows protocol to build an estimate for the second-order Renyi entropy where the corresponding purity is estimated as
the median among $N_{g}$ measurement groups each containing $M_{g}$ Pauli measurements, i.e.,
$\hat{P}_{\text{Shadows}} = \text{median} \{ \hat{P}_{\text{Shadows}, 1}^{(M_{\text{g}},K)}, \dots, \hat{P}_{\text{Shadows}, N_{\text{g}}}^{(M_{\text{g}},K)}\}$ with 
\begin{widetext}
\begin{align}
    \hat{P}_{\text{Shadows}}^{(M_{\text{g}},K)} &= \frac{2}{M_{\text{g}}(M_{\text{g}}-1)} \sum_{m} \sum_{m'| m'< m} \frac{1}{K^2} \sum_{k,k'=1}^{K} \prod_{n=1}^{N} \Big(\Big. 9\times \gamma_{n,m,m',k,k'} - 4\Big.\Big) \,,\label{eq:explicit-purity-estimate-Pauli-shadows}
\end{align}
\end{widetext}
where $n$ labels the qubits,  $k,k'$ the snapshot of a given measurement $m$, and 
$\gamma_{n,m,m',k,k'}\in\{0,1/2,1\}$ depends on the measurement and its outcome as
given in Appendix  \ref{app:purity_estimate_from_CS} in the case of Pauli measurements, and 
the total number of shadows is $M = N_{\text{g}} M_{\text{g}}$. 
Because the operator $S$ is not local, the use of classical shadows to estimate purities requires an exponential scaling of the number of samples with system size.

\begin{figure}[h!]
    \centering
    \includegraphics[width=.5\textwidth]{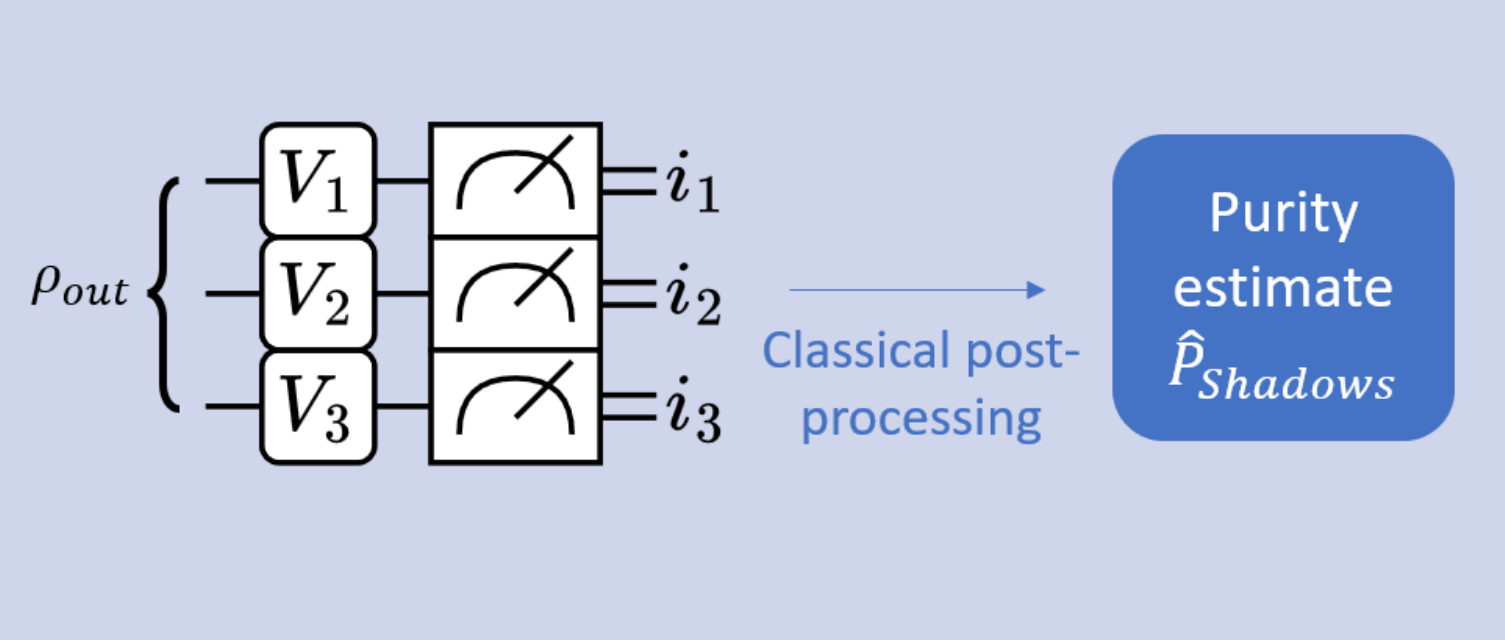}
    \caption{\textbf{Classical shadows protocol.} Each $V_i$ is a single-qubit Clifford gate such that its combination with a computational basis measurement corresponds to a random measurement in the $X$, $Y$, or $Z$ basis.}
    \label{subfig:CS_protocol}
\end{figure}

Although the Renyi-2 entropy or equivalently purity is a global property, here we choose to consider Pauli basis measurements for entropy estimation on \textit{near-term} devices. This is because a single layer of single-qubit gates is easy to implement and introduces fewer errors compared to an $n$-qubit Clifford circuit. Fig.~\ref{subfig:CS_protocol} shows the corresponding measurement circuit. For large quantum systems, the Pauli version of the method is very inefficient in terms of sampling cost, as the purity is a global property thus implying an $O(4^n)$ scaling with the system size $n$: fixing $\epsilon, \delta > 0$, obtaining an $\epsilon-$accurate purity estimate with success probability at least $1-\delta$ requires $O(\ln(\frac{1}{\delta}) \times \frac{4^n}{\epsilon^2})$
(Pauli) measurements (see Appendix \ref{app:purity_estimate_from_CS}). Note that the $n$-qubit Clifford version of the protocol for purity estimation would only provide a quadratic improvement of the scaling, as the number of measurements scales as $\sqrt{\Tr[S^2]} =2^n$. We expect that the added complexity of the circuit will make this approach unpractical for sizes beyond tens of qubits.

\subsection{Study of entropy accumulation on a QPU}\label{subsubsec:experimental_study_of_entropy_accumulation}

Using the Pauli version of the classical shadows protocol whose estimate is given in Eq.~\eqref{eq:explicit-purity-estimate-Pauli-shadows}, we estimate the purity and Renyi-2 entropy density of the quantum state along the chosen VQA circuit, i.e., as functions of the circuit depth, on Rigetti's Aspen-M-3 QPU (see Appendix \ref{app:calibration_data_Aspen-M-3} for the device's description). QPU results for a circuit width $n = 3$ can be found in Fig.~\ref{subfig:QPU_run_n=3-R2d}, using two different estimation strategies (see discussion in Sec. \ref{subsubsec:randomization_measurements_CS} for details) corresponding to light and dark gold error bars, where, for each instance, a purity estimate was calculated by taking $K = 1000$ classical snapshots of the output state for each of the $M = 5\times 4^3 = 320$ measurement settings considered, then taking the median of means over five groups. Before we discuss a comparison of the QPU data to a global depolarizing model and its later refinement with additional $T_1$ relaxation, we explain the reasoning behind having two sets of QPU data in Figs.~{\ref{subfig:QPU_run_n=3-R2d}} and \ref{subfig:QPU_run_n=3-R2d-T1}.

\subsubsection{Randomization of measurements}\label{subsubsec:randomization_measurements_CS}

A standard classical shadows protocol should ideally randomly select a new measurement for every measurement sample. Changing circuits being relatively costly in execution time for some platforms like superconducting qubits, we often implement a variant where $M$ measurement settings are randomly selected and each one is executed $K$ times its number of occurrence. Because our QPU data were limited in number of qubits ($n=3$) and therefore we were expecting to see the occurrence of most measurements and in similar number, we initially opted to derandomize the measurement choice and scan the $3^3=27$ possible measurement settings with $K=1000$ classical snapshots of the output state for each of them~\footnote{The original protocol~\cite{huang_predicting_2020} takes a single shot per random setting ($K=1$), but on platforms where reconfiguring the measurement circuit dominates the runtime, such as the superconducting QPU used here, a multishot variant ($K>1$) is more time efficient~\cite{elben_randomized_2023, zeng2023tailoring}. Reported values of $K$ in the literature range from $1$ \cite{zhang2021experimental} to $10^4-10^5$ \cite{struchalin2021experimental}. In this work, we use $K=1000$ for the hardware experiments, and take the same value for simulations to make a fair comparison. Although analyzed for specific observables \cite{zhou2023performance}, we note that the optimal split between the number of settings, $M$, and shots per setting, $K$, for purity estimation via Pauli-basis shadows is left as an open question.}.
We were surprised to observe that this strategy generated a systematic error in the QPU entropy estimation that was later corroborated by numerical simulations (see Appendix \ref{subsec:systematic_error_derand_CS}).

To mitigate this bias, we reintroduced a random choice of measurements in our QPU results (light and dark gold error bars in Fig.~\ref{fig:QPU_run_n=3_classim_and_QPU_CS}) through two different methods. 
We first select $M=320$ random measurements and count how many times each measurement is chosen. In method 1 (light gold error bars), for a fixed run of the protocol and a given choice of measurement, we build the list of shadows by replicating the measurement outcomes associated with that measurement from the corresponding derandomized run the given number of times the measurement setting was obtained. Since we still have three sets of data, we can obtain errors bars, exactly as done before.
In method 2 (dark gold crosses), we sacrifice the information on error bars in order to cluster the data of the previous three runs into a single one that contains more realistic randomized QPU data; i.e., we build the purity estimate for a single run of the protocol by exploiting the three available sets of snapshots for each measurement setting. More precisely, for a given setting, the first three times a measurement occurs, we use one of the three sets of data, and if a measurement occurs more times, we cycle over them again.

\subsubsection{Gap with respect to the global depolarizing model}\label{subsubsec:gap with depol}

\begin{figure*}
     \centering
     \begin{subfigure}[b]{\textwidth}
         \centering
         \includegraphics[width=\textwidth]{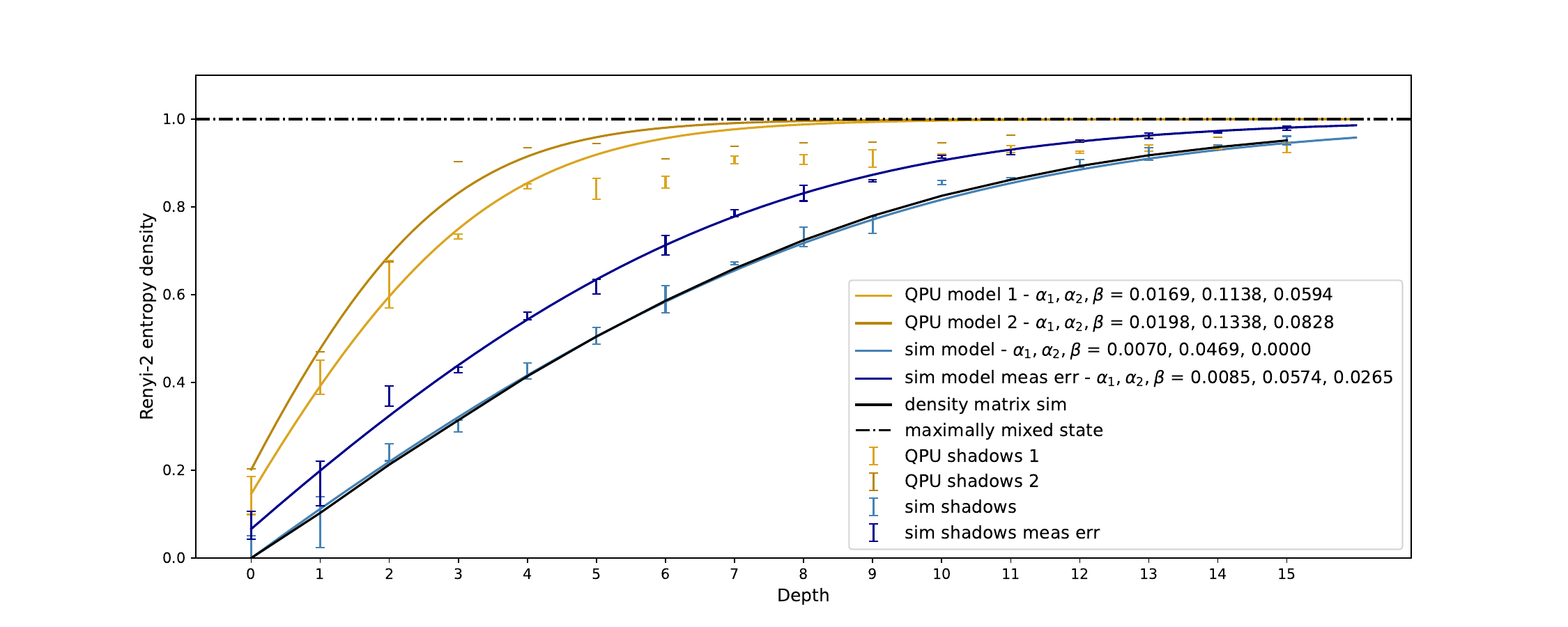}
         \caption{Without $T_1$ errors}
         \label{subfig:QPU_run_n=3-R2d}
     \end{subfigure}
     \begin{subfigure}[b]{\textwidth}
         \centering
         \includegraphics[width=\textwidth]{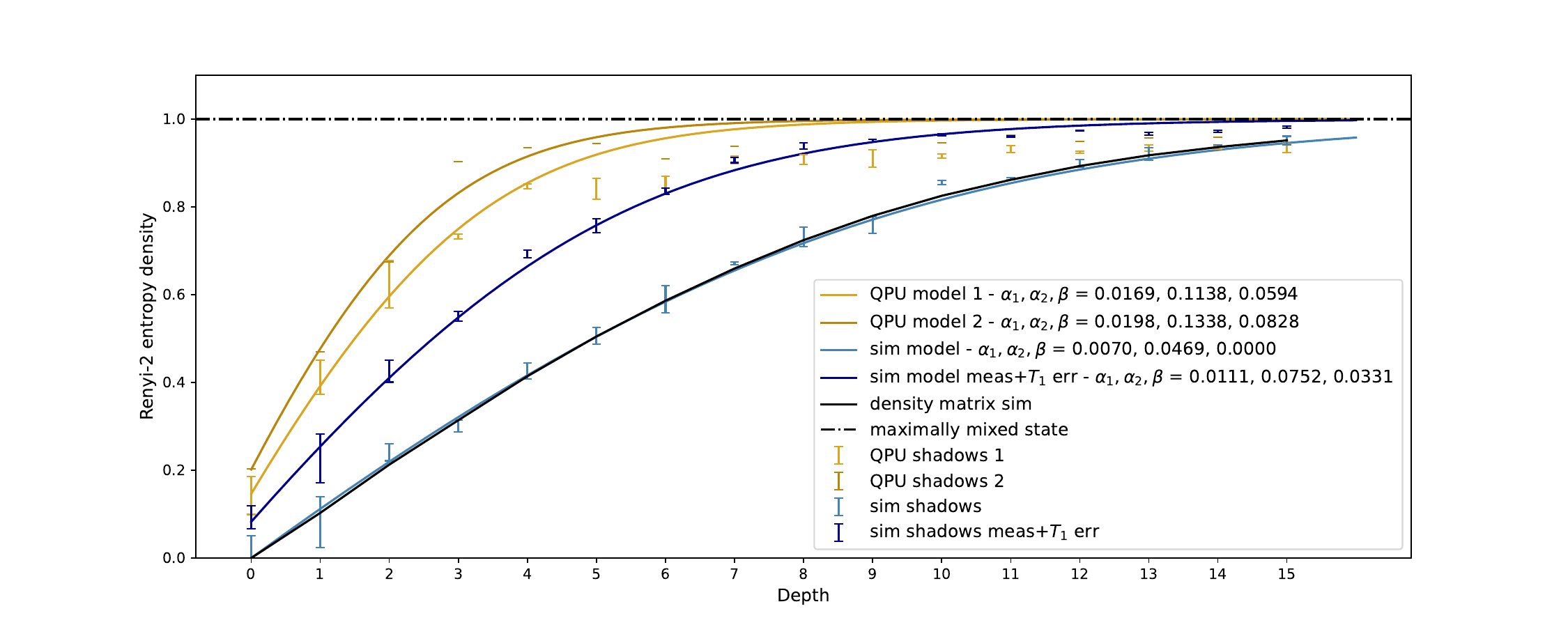}
         \caption{With $T_1$ errors}
         \label{subfig:QPU_run_n=3-R2d-T1}
     \end{subfigure}
        \caption{\textbf{QPU results and validation of our heuristic model.} We consider a VQA circuit as in Fig.~\ref{fig:ideal_hardware_efficient_param_circuit} applied to $n=3$ qubits, with fixed circuit parameters (fixed random seed \texttt{np.random.seed(837)}). Gold error bars give the purity and Renyi-2 entropy evolution as functions of circuit depth on Rigetti's QPU Aspen-M-3 using two different methods. Each error bar was obtained by running the classical shadows protocol three times, and computing the average estimate and the standard deviation over those three samples. Results obtained in the same setting but using a local depolarizing noise model ($p_1=0.008$ and $p_2=0.054$ from calibration data of Aspen-M-3) and without measurement errors correspond to the light blue error bars. The dark blue error bars in Fig.~\ref{subfig:QPU_run_n=3-R2d} correspond to the same simulation with added measurement errors from calibration data (measurement circuit gate errors $p_1=0.008$ and errors in the detectors $P(0|1)=0.03$ and $P(1|0)=0.02$). The dark blue error bars in Fig.~\ref{subfig:QPU_run_n=3-R2d-T1} also include $T_1$ relaxation errors both at the gate level ($\gamma_1, \gamma_2 = 0.003, 0.015$) right after depolarization ($p_1, p_2 = 0.006, 0.038$) and in the measurement stage ($\gamma_{meas}=0.05$) just before readout errors ($P(0|1)=P(1|0)=0.03$) based on calibration data. The black solid lines show the purity and Renyi-2 entropy density evolution obtained via density matrix simulation under our local depolarizing noise model without measurement errors (similar to the light blue error bars). The black horizontal dash-dotted lines correspond to the purity or entropy density of the maximally mixed state $\sigma_0\coloneqq I/2^n$. Fit of our global depolarizing heuristic model, where we have imposed that $\alpha_2/\alpha_1=p_2/p_1$, and where $\beta$ models readout errors and noise in the measurement circuit, corresponds to the solid gold and blue lines.}
        \label{fig:QPU_run_n=3_classim_and_QPU_CS}
\end{figure*}

Fig.~\ref{subfig:QPU_run_n=3-R2d} includes a comparison of the hardware entropy results with a density matrix simulation of the same VQA circuit under a local depolarizing noise model (see solid black line) with noise parameters $p_1 = 0.008$ and $p_2 = 0.054$ deduced from calibration data of Aspen-M-3 (see Appendix \ref{app:calibration_data_Aspen-M-3}). The QPU data show a faster-than-expected increase of the entropy density after only a few layers of gates when compared to a global depolarizing model with parameters resulting from calibration data. We also observe convergence to a value strictly below the threshold of $1$ corresponding to the maximally mixed state (dash-dotted line) with some minor oscillations for intermediate time. The nonzero entropy density at depth zero most probably indicates the presence of errors in the entropy estimation protocol.

One observes a clear gap between the QPU data and the density matrix simulation under local depolarizing noise, which can be easily interpolated with a global depolarizing model (light blue solid line), as done in Sec. \ref{subsubsec:analytical-model}.

To discard the possibility of the observed gap being solely a systematic error resulting from the classical shadows protocol, we perform a simulation of the target circuit under local depolarizing noise from calibration data followed by its noiseless classical shadows postprocessing under the same sample constraints as for the QPU data (see light blue error bars in Fig.~\ref{subfig:QPU_run_n=3-R2d}). We confirm that classical shadows do not seem to produce any significant systematic shift. Even if a very minor systematic error could be observed, it can be removed by further increasing the number of samples, as discussed in Appendix \ref{subsec:systematic_error_low_M_CS}.

To explain part of the observed gap between the real QPU data and the model, we incorporate errors in the gates of the measurement circuit (cf. Fig.~\ref{subfig:CS_protocol}) and readout errors into our numerical simulations, both deduced from calibration data (see Appendix \ref{app:calibration_data_Aspen-M-3}). The numerical results correspond to the dark blue error bars in Fig.~\ref{subfig:QPU_run_n=3-R2d}. To account for the additional measurement circuit gate error and additional readout errors, which may artificially increase the entropy estimation, we introduce an additional fitting parameter $\beta$ to our global depolarizing model from Eq.~\eqref{eq:purity-model-from-class-sim} (see Appendix \ref{app:renyi_entropy_density_model_CS}). The resulting fitted model corresponds to the dark blue solid line in Fig.~\ref{subfig:QPU_run_n=3-R2d}. We observe that its fitted parameters $\alpha_1$, $\alpha_2$, and $\beta$ closely match the calibration data $p_1=0.008$, $p_2=0.054$, and $p_m\approx P(0|1)=0.03$ used in the local depolarizing noise simulations with measurement circuit and readout errors (dark blue error bars). Although the added measurement circuit and readout noise account for part of the gap between the real hardware data and the depolarizing noise model, a significant part remains unexplained.

\subsubsection{Refinement with $T_1$ relaxation}
\label{Sec Exp - Discussion results}

One can notice in Fig.~\ref{subfig:QPU_run_n=3-R2d} that the heuristic global depolarizing noise model (dark blue solid line), despite capturing part of the QPU entropy accumulation behavior, also significantly differs. We observe three different features in the hardware results (gold error bars): (1) faster initial entropy growth, (2) potential saturation below the maximum entropy density (value of 1)  in the large-depth limit (confirmed by additional QPU data for circuits with up to depth 30), and (3) potential oscillations in the intermediate regime. As discussed previously, neither measurement errors nor a deviation of the classical shadows estimate is the reason for the observed differences.

Noticing the lengthy measurement time of the device and in an attempt to account for the above three features, Fig.~\ref{subfig:QPU_run_n=3-R2d-T1} shows an additional numerical simulation (dark blue error bars) that includes the effect of $T_1$ relaxation, modeled as amplitude damping channels, both in the circuit implementation and in the measurement stage (from calibration data of the device; see Appendix \ref{app:calibration_data_Aspen-M-3}). We emphasize that the local depolarizing probabilities used in our simulations are fixed from the single- and two-qubit gate fidelities reported in the device calibration data from which we removed a $T_1$ contribution. These are thus effective error rates that group various error mechanisms degrading the gates, including the often significant dephasing ($T_2$) contribution. We single out only $T_1$ relaxation for separate treatment, modeling it with an amplitude damping channel, as a purely depolarizing description cannot reproduce two qualitative features of the measured entropy accumulation: its saturation below the maximally mixed value and the oscillations seen at large depth. We do not introduce a separate dephasing channel, for two reasons. First, we aim to keep the effective noise model as simple as possible. Second, the random single-qubit rotations of the hardware-efficient ansatz are expected to twirl dephasing errors toward an isotropic, depolarizinglike channel \cite{wallman2016noise}, so that their effect on entropy is well captured by the effective local depolarizing channel. This also supports the applicability of the global depolarizing heuristic underlying our analysis.

The addition of $T_1$ relaxation into the simulation explains a significant part of the fast entropy growth. Moreover, the saturation of the entropy density below the maximally mixed state threshold of $1$, along with the observed oscillatory behavior, is partially accounted for by the added relaxation noise. However, a gap between the global depolarizing prediction (dark blue solid line) and the QPU data (gold error bars) remains. We believe that this could result from a more complex combination of errors present on the device, including coherent errors, crosstalk, or some additional damping not captured in our model, simulations, or in the original calibration data.

A natural direction for future work could be to incorporate a small number of additional effective noise parameters to model these other sources of error. 
To preserve the simplicity of the heuristic model, this could be done in a coarse-grained way, for example, by including nearest-neighbor correlated Pauli noise to capture crosstalk, or local coherent noise to model systematic overrotations. 
We note that while coherent unitary errors do not directly change the entropy they may interact with nonunital noise, such as amplitude damping, and thereby affect the entropy dynamics. 
Including such additional noise features would retain the analytical simplicity needed for circuit-volume estimates while more accurately capturing entropy saturation and oscillatory features in the QPU data.

\subsubsection{Discussion on the update of the heuristic model}

Our initial goal was to build simple heuristic models and prove their validity so they could later be exploited to make predictions at the application level. Therefore, a heuristic model that better captures the faster entropy increase would provide more stringent and accurate predictions on the performance of QPU hardware. Unfortunately, to the best of our knowledge, building a simple analytical model that captures this faster entropy increase resulting from the interplay of damping and depolarization is a nontrivial theoretical result.

While we wait for better models, we believe that a global depolarizing model built from calibration data provides a convenient compromise for the quantum device considered. Indeed, it is important to point out that such a model remains a lower bound to the QPU data for the most relevant part of the curve. This means that any prediction of effective available circuit volume done in the last step of the benchmarking methodology (presented in the next section) remains valid; it just becomes conservative and generous to the QPU performance. 

\section{\label{sec:App1_Quantum_Advantage_Benchmarking} Quantum Advantage Benchmarking}

After proposing a heuristic model, later tested and refined with numerical and hardware entropy accumulation analyses, we now present the last step of our benchmarking methodology. We will show how to combine the design of heuristic models of entropy accumulation in VQA circuits, such as the global depolarizing noise model of Eq.~\eqref{eq:purity-model-from-class-sim}, with the application-level quantum advantage benchmarking technique introduced in Ref. \cite{stilck_franca_limitations_2021}. This allows to make quick reliable predictions on whether the accumulation of errors precludes from reaching quantum advantage or the latter remains possible. Most importantly, one does not need to run the quantum processing unit to solve the targeted application; only a model of how entropy accumulates on the problem-specific circuit ansatz is required. In what follows, we revisit the analysis of Ref. \cite{stilck_franca_limitations_2021} for Max-Cut, but the technique is widely translatable to other optimization problems, whether defined by classical (e.g., combinatorial optimization problems \cite{perez2024variational}) or quantum Hamiltonians (e.g., quantum chemistry problems \cite{o2016scalable, stanisic2022observing}). 

\subsection{Background}

As detailed in Ref. \cite{stilck_franca_limitations_2021}, given a quantum device and known state-of-the-art classical solver solution to Max-Cut, we can certify whether there is some potential for a QPU to outperform the best classical solvers or conversely that quantum advantage is out of reach. 

\begin{figure}[h]
    \centering
    \includegraphics[width=0.5\textwidth]{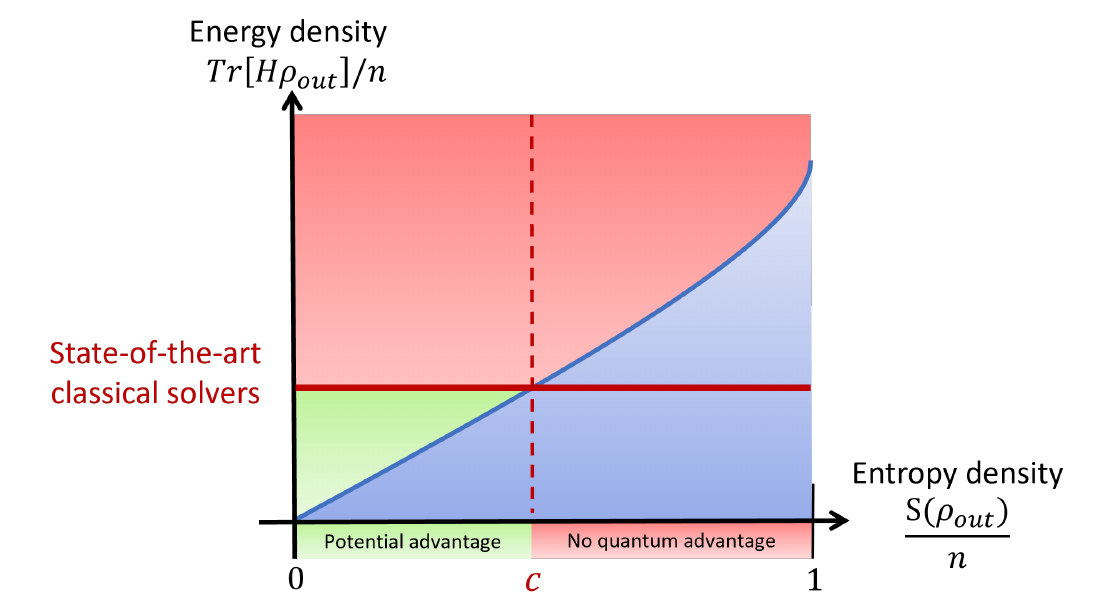}
    \caption{\textbf{Quantum advantage benchmarking framework from Ref. \cite{stilck_franca_limitations_2021}.} For an optimization task of the form $\min_{\rho} Tr[H\rho]$, the authors derive a lower bound (blue solid line) on the quantum device's solution to the energy density ($\Tr[H\rho_{out}]/n$) as a function of the accumulated entropy density ($S(\rho_{out})/n$). The state-of-the-art classical solvers' solution to this optimization problem (red horizontal line) defines an entropy density threshold $c$ such that if the quantum device's candidate solution $\rho_{out}$ satisfies $\frac{S(\rho_{out})}{n} \geq c$ then quantum advantage is out of reach for this device and for this optimization problem.}
    \label{fig:QAdvBenchTechnique-Stilck_Franca_Garcia-Patron}
\end{figure}

This is ascertained by an approach summarized in Fig.~\ref{fig:QAdvBenchTechnique-Stilck_Franca_Garcia-Patron}. Combining a lower bound on the quantum device energy density solution $\Tr[H\rho_{out}]/n$ as a function of the von Neumann entropy $S(\rho_{out})/n$ (blue solid line) with the classical solver energy density prediction (red horizontal line)
defines an entropy density threshold $S(\rho_{out})/n\geq c$ (red dotted vertical line) beyond which the QPU cannot provide a better solution than the classical solver.

Despite the benchmarking approach of Ref. \cite{stilck_franca_limitations_2021} being constructed for von Neumann entropies and our gathered QPU data being for the Renyi-2 entropy,
the well-known bound between the von Neumann and the Renyi entropies, $S(\rho) \geq S^{(2)}(\rho)\,, \forall \rho$, implies that any conclusion obtained replacing $S(\rho)$ by  $S^{(2)}(\rho)$ in the analysis summarized in Fig.~\ref{fig:QAdvBenchTechnique-Stilck_Franca_Garcia-Patron} is a conservative and valid prediction. 

\subsection{Bounds on quantum advantage via circuit-size bounds}
The next step of the approach is to combine the previous entropic threshold with the selected heuristic model of entropy accumulation in order to find a bound on the number of gates or maximum circuit depth before our QPU of interest can be certified to not provide better solutions than a standard classical solver. In what follows, we revisit the analysis for Max-Cut in Ref. \cite{stilck_franca_limitations_2021} but now equipped with heuristic models of entropy accumulation.

\subsubsection{Circuit bounds}

Now using a global depolarizing heuristic model for the second-order Renyi entropy of the output of the VQA ansatz described in Eq.~\eqref{eq:purity-model-from-class-sim}, we obtain
the following bound on the circuit size (with width $n$ and depth $D$):
\begin{equation}
    2(2\alpha_1 n + \alpha_2 (n - 1)) D\geq \ln \left( \frac{2^n - 1}{2^{n(1-c)} - 1} \right) \,.
\end{equation}

Assuming that 2-qubit gate depolarizing noise is the dominant source of noise on the device ($\alpha_2 = p_2 \neq 0$), which is generally the case in practice, and thus neglecting single-qubit gate depolarizing noise ($\alpha_1 = 0$), this yields

\begin{align}
     (n - 1) D & \geq \frac{1}{2  p_2} \ln \left( \frac{2^n - 1}{2^{n(1-c)} - 1} \right)\,,\label{eq:class_superiority_condition_2Q_dep_noise_only}
\end{align}
where we assume that $p_2$ is below $1\%$, a reasonable assumption for current QPUs. For large problem size $n\rightarrow \infty$, this leads to the bound on the circuit depth (number of circuit layers):
\begin{align}
      D &\geq \frac{1}{2p_2} c\ln \left( 2 \right)\,.\label{eq:class_superiority_condition_2Q-dep_noise_only_large_n}
\end{align}

Note the similarity with the depth threshold obtained in Eq.~\eqref{eq:depth_threshold_maximised_entropy_large_systems} for large systems $n\rightarrow \infty$. In the special case where two-qubit gate depolarizing noise dominates ($\alpha_1 = 0$, $\alpha_2 = p_2$), both depth thresholds differ by the scalar factor $c$ resulting from the entropy density threshold analysis above.

\subsubsection{Comparison with previous work}

Gset is a set of instances of Max-Cut popular for performance benchmarking for which classical solvers' solutions and times to solution are known. Following Ref. \cite{stilck_franca_limitations_2021}, we use the G12 instance of Gset consisting of $800$ nodes and $1600$ edges. Using Eq.~\eqref{eq:class_superiority_condition_2Q_dep_noise_only} and the plot of the bound for this Max-Cut instance from Ref. \cite{stilck_franca_limitations_2021} where the entropy density threshold against a heuristic classical solver reads $c = 0.3$, we conclude that for $D \geq 0.104/p_2$, quantum advantage is out of reach for a quantum device with two-qubit gate depolarizing probability $p_2$; as we expect, this is an even smaller depth threshold than the $D \geq 0.18/p_2$ predicted in Ref. \cite{stilck_franca_limitations_2021}. For a current state-of-the-art two-qubit gate error of $p_2=10^{-3}$ \cite{acharya2024quantum}, our approach predicts that circuits with depth $D\geq 110$ are already in the regime of certified classical superiority.

\begin{figure}[h!]
    \centering
    \includegraphics[width=.48\textwidth]{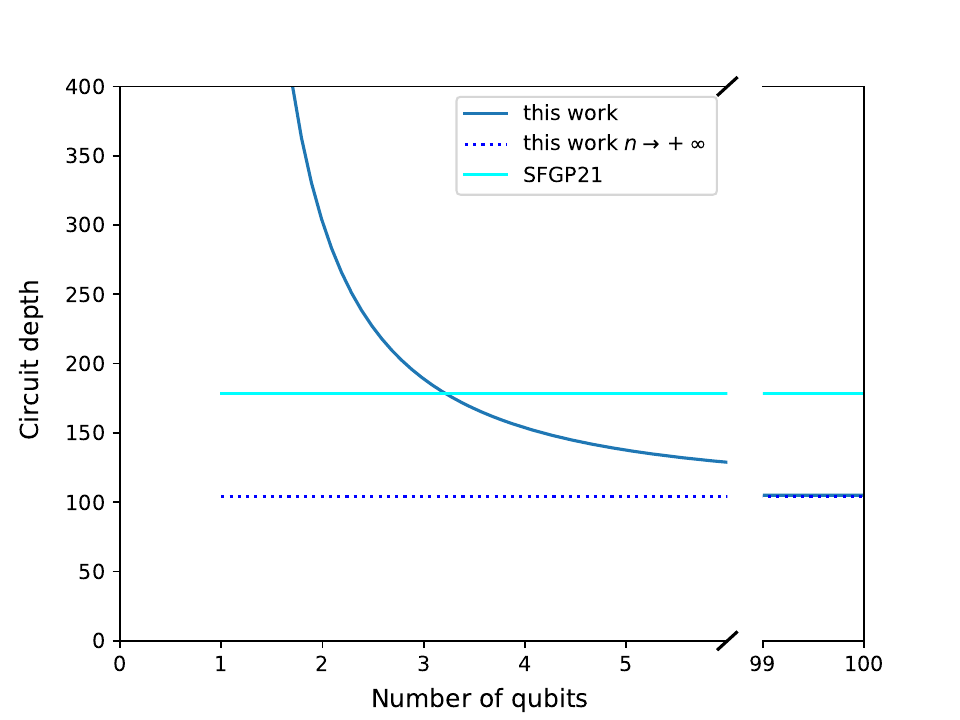}
    \caption{\textbf{Circuit size threshold above which quantum advantage is out of reach} (for arbitrary $c$ and $p_2$ values; here $c=0.3$ and $p_2=10^{-3}$). `This work' corresponds to the newly derived circuit-size bound obtained from Eq.~\eqref{eq:class_superiority_condition_2Q_dep_noise_only} (dark blue solid line)  while SFGP21 (light blue solid line) is the depth condition previously derived in Ref. \cite{stilck_franca_limitations_2021}. We cut the x axis to show the threshold behavior in the $n \rightarrow \infty$ limit.}
    \label{fig:circuit_size_qadv_boundary}
\end{figure}

In a more general setting, our approach also allows to predict an effective circuit volume, i.e., total number of gates obtained from Eq.~\eqref{eq:class_superiority_condition_2Q_dep_noise_only},
beyond which quantum advantage for the specific problem under study is no longer possible. The effective circuit volume is visualized in Fig.~\ref{fig:circuit_size_qadv_boundary} (dark blue line) for the G12 Max-Cut instance above for a QPU with two-qubit gate error given by $p_2=10^{-3}$.

For large system sizes, we see that our depth threshold saturates at a constant value $\frac{1}{2p_2} c\ln \left( 2 \right)$ for fixed $c$ (dotted blue line), which is below the depth threshold derived in Ref. \cite{stilck_franca_limitations_2021} based on relative entropy contraction results (cyan line). In Fig.~6, keeping $c=0.3$ fixed as $n$ increases is a simplifying assumption used to illustrate the large system behavior of our circuit-size bound and should not be interpreted as implying that $c$ is generally independent of the problem size. This shows that quantum advantage is lost at a shorter depth than originally predicted for large problem sizes. In fact, it can be proven analytically that our depth threshold is tighter than that of Ref.  \cite{stilck_franca_limitations_2021} for any $c$ and $p_2$ values (see Appendix \ref{app:circuit_size_boundary_condition_for_qadv}). One can also prove that in the event of the threshold depth still keeping some dependency on the number of qubits, $n$, our condition in Eq.~\eqref{eq:class_superiority_condition_2Q-dep_noise_only_large_n} would still be tighter than the depth threshold from Ref. \cite{stilck_franca_limitations_2021} for large systems (see Appendix \ref{app:circuit_size_boundary_condition_for_qadv}).

\section{\label{sec:Conclusion}Discussion}

Inspired by the theoretical framework of Ref.  \cite{stilck_franca_limitations_2021}, we have presented a QPU benchmarking methodology using the accumulation of entropy along the computation as a performance metric, which allows to bridge between circuit-level and application-level benchmarking techniques. The method works by first proposing simple but sufficiently explanatory models of entropy accumulation, and then by validating and refining those models via comparison to QPU data. These models are then used to estimate the available volume of computation before any potential quantum advantage is lost for our optimization problem of interest. We have carried out a proof-of-principle demonstration of the methodology centered on the global depolarizing model and, after comparison with QPU data, we have proposed some refinements accounting for the effect of $T_1$ relaxation time.
Using this simple model for entropy accumulation, we have shown how to impose more stringent bounds than previously shown in Ref. \cite{stilck_franca_limitations_2021} on the circuit size above which quantum advantage is lost for hardware-efficient VQA circuits run on a typical superconducting NISQ device.

When estimating circuit purities and their corresponding entropies, any imperfection in the measurement process---whether it is a noisy gate in the measurement protocol, $T_1$ relaxation, or bit flips occuring during the measurement---can cause the final entropy estimate to deviate from the target true circuit output entropy. Therefore, further work is needed to develop error mitigation techniques specific for purity and entropy estimation, for example, by adapting existing SPAM error mitigation techniques to the peculiarities of each estimation protocol. Another interesting direction would be to develop theoretical techniques that would provide guarantees for the outcome of a given, potentially error-mitigated, entropy estimation protocol.

For medium and large systems, we expect classical shadows to suffer from larger error bars and potentially stronger systematic errors, making it unpractical, leaving the SWAP test as the only known viable option. If sample-efficient single-copy techniques should not exist, following complexity arguments \cite{acharya2020estimating, chen2022exponential}, an interesting open question is whether some relaxation of the problem, such as restricting to upper and/or lower bounds, leads to sample-efficient protocols.

Although depolarizing noise seems to be a decent first approximation model to lower bound entropy accumulation in superconducting circuits run on a typical NISQ device (here a two-qubit gate fidelity of the order of $10^{-1}$), our results hint that a more complete model for entropy accumulation is required to make tighter predictions for such quantum devices, for example, incorporating amplitude damping, coherent errors, and crosstalk. A challenge here is to have simple theoretical models, such as the global depolarization, that capture most of these additional features. From our experience, the interplay of amplitude damping and depolarizing seems to lead to nontrivial dynamics and regimes that require a more detailed investigation to build more powerful models.

As superconducting quantum computers scale up and noise levels are reduced, we expect, inspired by Ref. \cite{dalzell2024random}, that the global depolarizing heuristic model should provide a better approximation to the entropy accumulation behavior under local depolarizing noise, which in turn would become a better assumption to the real QPU entropy signature as the effect on entropy of other sources of errors such as amplitude damping become negligible. We leave the confirmation of this trend as a very relevant open question.

We expect our methodology to carry over to platforms beyond superconducting systems, and to topologies other than nearest-neighbor planar superconducting circuit architectures studied in this work, with the possible exception of photonic platforms in the pre-fault-tolerant regime. This is because photon loss, a dominant source of imperfection in photonic platforms, has the vacuum state as its fixed point, which has zero entropy. Therefore, our entropy metric will need to be complemented by further performance metrics for pre-fault-tolerant photonic devices. It is important to point out that our methodology can also be applied to quantum error-corrected and fault-tolerant architectures by focusing on the entropy of the logical qubits. In a fault-tolerant regime, we expect the physical platform and architecture to become less relevant. Interestingly, the validity of our methodology could be restored for photonic platforms in the fault-tolerant regime, as such architectures are now meant to encode logical qubits being affected by standard qubit noise, which we expect to mostly increase the entropy. There are two open questions related to our methodology and the nature of the platform that are worth further investigation. The first is whether the accumulation of entropy density as the circuit depth increases has any dependency on the type of platform and its topology. On the one hand, our simulations seem to indicate that a global depolarizing model becomes more accurate as the gate quality improves (at least for the one-dimensional architecture studied here); on the other hand, an all-to-all connectivity architecture allows entanglement to spread faster and therefore a potentially stronger sensitivity to local imperfections, leading to a potentially faster increase in entropy. Settling this question is an interesting direction for future research. The second is whether the existence of simple models of entropy density accumulation (here, a global depolarizing model was a first approximation) depends on the choice of platform and its architecture topology in the pre-fault-tolerant era. While this is certainly true for photonic platforms, comparing ions and neutral atoms to superconducting circuits is an interesting question for further investigation.

\begin{acknowledgments}
The authors thank Pauline Besserve for useful feedback on this paper, and Marco Paini, Sean Thrasher, Chirag Wadhwa, and Qisheng Wang for valuable discussions. 
Results on the QPU were obtained using Rigetti’s Aspen-M-3 system as part of the Innovate UK Project No. 44167: Quantum Computing Platform for NISQ Era Commercial Applications. R.G.-P. and K.H. were supported by the EPSRC-funded project Benchmarking Quantum Advantage, Reference No. EP/Y004418/1.
\end{acknowledgments}

\section*{Data availability}
The data that support the findings of this article are openly available \cite{Demarty_EntropyBenchmarking}.

\bibliography{Entropy_Benchmarking_Paper}

\onecolumngrid

\newpage

\appendix

\input{app-Heuristic_model_CS_measerr}

\input{app-CS_for_purity}

\input{app-QAdv_benchmarking}

\input{app-Aspen-M-3-calibration-data}

\end{document}

%% file: app-Heuristic_model_CS_measerr.tex
\section{Analytical model for the Renyi entropy density evolution in a VQA circuit with classical shadows and readout noise}\label{app:renyi_entropy_density_model_CS}

In this section, we adapt the heuristic model from Eq.~\eqref{eq:purity-model-from-class-sim} to take into account the effect of noise in the measurement circuit of the classical shadows protocol and readout errors on the Renyi-2 entropy and purity evolution.

We assume that we can represent local depolarizing noise, and readout and measurement circuit noise, by two successive global depolarizing noise channels $\mathcal{N}_P : \rho \mapsto (1-P)\rho + P \frac{I}{2^n}$ and $\mathcal{N}_{P_M} : \rho \mapsto (1-P_M)\rho + P_M \frac{I}{2^n}$ (for an $n$-qubit quantum state $\rho$) with depolarizing probabilities $P$ and $P_M$, respectively, placed at the very end of the ideal circuit. As we did in Eq. \ref{eq:link-global-noise-param-and-fitting-params}, we can link those two probabilities to fitting parameters $\alpha_1, \alpha_2$ (depolarizing) and $\beta$ (readout) by equating the probabilities of no error occurring in both our local noise model and global noise model:
\begin{align}
    1 - P &\coloneqq (1 - p_1)^{|\text{1Q gates}|} (1 - p_2)^{|\text{2Q gates}|}\,,\\
    &= (1 - p_1)^{2nD} (1 - p_2)^{(n-1)D}\,,\\
    &\coloneqq e^{-2\alpha_1 nD} e^{-\alpha_2 (n-1)D}\,,
\end{align}
and
\begin{equation}
    1 - P_M \coloneqq (1 - p_m)^n \coloneqq e^{-\beta n}\,,
\end{equation}
where $\alpha_i = \ln((1-p_i)^{-1})$ and $\beta = \ln((1-p_m)^{-1})$ are the local depolarizing error probabilities $p_i$ and $p_m$ in the low-noise limit, $\alpha_i \approx p_i$ and $\beta \approx p_m$.
Our model for the purity is given by
\begin{equation}
    \Tr[\rho_{D}^2] = (1 - 2^{-n})(e^{-2(2\alpha_1 nD +\alpha_2(n-1)D + \beta n)} - 1) + 1\,.\label{eq:purity_model_CS_measerr}
\end{equation}

%% file: app-CS_for_purity.tex
\section{Purity estimate from the classical
shadows protocol with Pauli basis
measurements}\label{app:purity_estimate_from_CS}

In this section dedicated to purity estimation using the Pauli-basis classical shadows protocol from Ref. \cite{huang_predicting_2020}, we provide a simple expression for the corresponding purity estimate in Appendix \ref{subsec:purity_estimate}, give an upper bound on the sampling cost in Appendix \ref{subsec:upper_bound_sampling_cost}, analyze the performance of the protocol based on how mixed the target state is in Appendix \ref{subsec:variance-bound-depends-on-purity-CS-protocol}, and discuss the existence of a systematic error in the purity estimate of a derandomized version of the classical shadows technique in Appendix \ref{subsec:systematic_error_derand_CS}.

\subsection{Purity estimate}\label{subsec:purity_estimate}

Consider the purity estimate from the classical shadows protocol given in Ref. \cite{elben_randomized_2023}:
\begin{equation}
    \hat{P}_{\text{Shadows}}^{(M,K)} \coloneqq \frac{1}{M(M-1)} \sum_{m \neq m'} Tr[\hat{\rho}^{(m)}\hat{\rho}^{(m')}]\,,\label{eq:CS-purity-estimate-vs-shadows}
\end{equation}
where 
\begin{equation}
    \hat{\rho}^{(m)} \coloneqq \frac{1}{K} \sum_{k=1}^{K} \bigotimes_{n=1}^{N} \left( 3 \left(U_n^{(m)} \right)^\dagger \ket{s_n^{(m,k)}}\bra{s_n^{(m,k)}} U_n^{(m)} - I \right)\,.
\end{equation}

For the special case of Pauli measurements, it is possible to further simplify the formula for the purity estimate. First, consider the product $\hat{\rho}^{(m)}\hat{\rho}^{(m')}$:

\begin{align}
    \hat{\rho}^{(m)}\hat{\rho}^{(m')} &= \frac{1}{K^2} \sum_{k,k'=1}^{K} \bigotimes_{n=1}^{N} 
    \left( 3 \left(U_n^{(m)} \right)^\dagger \ketbra{s_n^{(m,k)}}{s_n^{(m,k)}} U_n^{(m)} - I \right) \nonumber\\
    &\quad \left( 3 \left(U_n^{(m')} \right)^\dagger \ketbra{s_n^{(m',k')}}{s_n^{(m',k')}} U_n^{(m')} - I \right) \,,\\
    &= \frac{1}{K^2} \sum_{k,k'=1}^{K} \bigotimes_{n=1}^{N} \\
    &\quad \Big(\Big. 9 \left(U_n^{(m)} \right)^\dagger \ketbra{s_n^{(m,k)}}{s_n^{(m,k)}} U_n^{(m)} \left(U_n^{(m')} \right)^\dagger \ketbra{s_n^{(m',k')}}{s_n^{(m',k')}} U_n^{(m')}\\
   &\quad - 3 \left(U_n^{(m)} \right)^\dagger \ketbra{s_n^{(m,k)}}{s_n^{(m,k)}} U_n^{(m)} - 3 \left(U_n^{(m')} \right)^\dagger \ketbra{s_n^{(m',k')}}{s_n^{(m',k')}} U_n^{(m')}+ I \Big.\Big) \,.
\end{align}

Taking the trace of the above product of shadows and exploiting the linearity of the trace and the fact that the trace of a tensor product is the product of the traces, we have

\begin{align}
    \Tr[\hat{\rho}^{(m)}\hat{\rho}^{(m')}] &= \frac{1}{K^2} \sum_{k,k'=1}^{K} \prod_{n=1}^{N} \\
    &\quad \Big(\Big. 9\Tr[ \left(U_n^{(m)} \right)^\dagger \ketbra{s_n^{(m,k)}}{s_n^{(m,k)}} U_n^{(m)} \left(U_n^{(m')} \right)^\dagger \ketbra{s_n^{(m',k')}}{s_n^{(m',k')}} U_n^{(m')}] \label{eq:app-b10}\\
   &\quad - 3\Tr[ \left(U_n^{(m)} \right)^\dagger \ketbra{s_n^{(m,k)}}{s_n^{(m,k)}} U_n^{(m)}]\\
    &\quad - 3\Tr[ \left(U_n^{(m')} \right)^\dagger \ketbra{s_n^{(m',k')}}{s_n^{(m',k')}} U_n^{(m')}]+ \Tr[I] \Big.\Big)\,,
\end{align}

where each of the traces above are traces of 2-by-2 matrices, so that
\begin{equation}
    \Tr[I] = 2\,.
\end{equation}
As the trace is cyclic and the $U_n^{(m)}$ are unitary matrices and since the $\ket{s_n^{(m',k')}}$ form an orthonormal family of quantum states, we deduce that
\begin{align}
    \Tr[ \left(U_n^{(m)} \right)^\dagger \ketbra{s_n^{(m,k)}}{s_n^{(m,k)}} U_n^{(m)}] &= 1\,,\\
    \Tr[ \left(U_n^{(m')} \right)^\dagger \ketbra{s_n^{(m',k')}}{s_n^{(m',k')}} U_n^{(m')}] &=1\,.
\end{align}
Simplifying \eqref{eq:app-b10} needs further work; however, for the special case of Pauli measurements, possible values for this trace can be restricted to the three values $\{0,1,1/2\}$. To see this, first notice that since the trace is cyclic,

\begin{align}
    &\Tr[ \left(U_n^{(m)} \right)^\dagger \ketbra{s_n^{(m,k)}}{s_n^{(m,k)}} U_n^{(m)} \left(U_n^{(m')} \right)^\dagger \ketbra{s_n^{(m',k')}}{s_n^{(m',k')}} U_n^{(m')}] \\
    \quad&= \Tr[ \bra{s_n^{(m,k)}} U_n^{(m)} \left(U_n^{(m')} \right)^\dagger \ketbra{s_n^{(m',k')}}{s_n^{(m',k')}} U_n^{(m')}\left(U_n^{(m)} \right)^\dagger \ket{s_n^{(m,k)}}]\,,\\
    \quad&= \Tr[\left|\bra{s_n^{(m,k)}} U_n^{(m)} \left(U_n^{(m')} \right)^\dagger \ket{s_n^{(m',k')}}\right|^2]\,,\\
    \quad&= \left|\bra{s_n^{(m,k)}} U_n^{(m)} \left(U_n^{(m')} \right)^\dagger \ket{s_n^{(m',k')}}\right|^2=\gamma_{n,m,m',k,k'}\,,
\end{align}

where $s_n^{(m,k)},s_n^{(m',k')} \in \{0,1\}$ and $U_n^{(m)}, U_n^{(m')} \in \{ H, HS^\dagger, I\}$. Exploring all possible combinations for $s_n^{(m,k)},s_n^{(m',k')}, U_n^{(m)}, U_n^{(m')} $, it can then be shown that $\gamma_{n,m,m',k,k'}$ can only take three different kinds of values: this is summarized in Table~\ref{tab:possible_values_norm_squared_complex_number_in_purity_estimate}.

\begin{table}[h]
    \centering
    \begin{tabular}{c|c|c|c|c}
        $U_n^{(m)} \times {\left(U_n^{(m')}\right)}^{\dagger}$ & $\left|\bra{0}.\ket{0}\right|^2$ & $\left|\bra{0}.\ket{1}\right|^2$ & $\left|\bra{1}.\ket{0}\right|^2$ & $\left|\bra{1}.\ket{1}\right|^2$\\ \hline
        $H \times H = I$ & 1 & 0 & 0 & 1\\
        $H \times SH = HSH$ & 1/2 & 1/2 & 1/2 & 1/2\\
        $H \times I = H$ & 1/2 & 1/2 & 1/2 & 1/2\\
        $HS^\dagger \times H = HS^\dagger H$ & 1/2 & 1/2 & 1/2 & 1/2\\
        $HS^\dagger \times SH = I$ & 1 & 0 & 0 & 1\\
        $HS^\dagger \times I = HS^\dagger$ & 1/2 & 1/2 & 1/2 & 1/2\\
        $I \times H = H$ & 1/2 & 1/2 & 1/2 & 1/2\\
        $I \times SH = SH$ & 1/2 & 1/2 & 1/2 & 1/2\\
        $I \times I = I$ & 1 & 0 & 0 & 1
    \end{tabular}
    \caption{Possible values for $\gamma_{n,m,m',k,k'}$ depending on the value of $s_n^{(m,k)},s_n^{(m',k')}, U_n^{(m)}, U_n^{(m')}$} where $\gamma_{n,m,m',k,k'} \coloneqq \left|\bra{s_n^{(m,k)}} U_n^{(m)} \left(U_n^{(m')} \right)^\dagger \ket{s_n^{(m',k')}}\right|^2$
    \label{tab:possible_values_norm_squared_complex_number_in_purity_estimate}
\end{table}

Finally, combining all previous results for the trace, we conclude that, for Pauli measurements, the purity estimate can be simplified as

\begin{equation}
    \hat{P}_{\text{Shadows}}^{(M,K)} = \frac{1}{M(M-1)} \sum_{m \neq m'} \frac{1}{K^2} \sum_{k,k'=1}^{K} \prod_{n=1}^{N} \Big(\Big. 9\times \gamma_{n,m,m',k,k'} - 4\Big.\Big)\,.
\end{equation}

Noting that $\gamma_{n,m,m',k,k'} = \gamma_{n,m',m,k,k'}$, this further simplifies to

\begin{equation}
\boxed{
     \hat{P}_{\text{Shadows}}^{(M,K)} = \frac{2}{M(M-1)} \sum_{m} \sum_{m'| m'< m} \frac{1}{K^2} \sum_{k,k'=1}^{K} \prod_{n=1}^{N} \Big(\Big. 9\times \gamma_{n,m,m',k,k'} - 4\Big.\Big)
    }\,,
\end{equation}
where 
\begin{equation}
    \gamma \coloneqq \gamma_{n,m,m',k,k'} = \left|\bra{s_n^{(m,k)}} U_n^{(m)} \left(U_n^{(m')} \right)^\dagger \ket{s_n^{(m',k')}}\right|^2\,,
\end{equation}
is given by Table~\ref{tab:possible_values_norm_squared_complex_number_in_purity_estimate}.

Considering $N_{\text{g}}$ groups containing $M_{\text{g}}$ measurement settings each, we may build an independent purity estimate for each group, then obtain the median of those purity estimates:
\begin{align}
\boxed{
    \hat{P}_{\text{Shadows}} = \text{median} \{ \hat{P}_{\text{Shadows}, 1}^{(M_{\text{g}},K)}, \dots, \hat{P}_{\text{Shadows}, N_{\text{g}}}^{(M_{\text{g}},K)}\}\,,
    }
\end{align}
where we set $M = N_{\text{g}} M_{\text{g}}$.

\subsection{Upper bound on the sampling cost}\label{subsec:upper_bound_sampling_cost}

Fix accuracy parameters $\delta, \epsilon > 0$ and $K=1$. 
Set $M = N_{\text{g}} M_{\text{g}}$, where
\begin{equation}
    N_{\text{g}} = 2\ln(2/\delta)\,,
\end{equation}
and
\begin{equation}
    M_{\text{g}} = \frac{272}{\epsilon^2} \times 4^n ||S||_\infty^2 = \frac{272}{\epsilon^2} \times 4^n = O(4^n / \epsilon^2)\,.
\end{equation}
From Theorem S4 and Proposition S4 of the Supplementary Material of Ref. \cite{huang_predicting_2020}, we know that a collection of $M=N_{\text{g}} M_{\text{g}}$ independent classical shadows allows one to accurately predict the purity $\Tr[\rho^2]$ of state $\rho$:
\begin{equation}
    |\hat{P}_{\text{Shadows}} - \Tr[\rho^2]| \leq \epsilon\,,
\end{equation}
with probability at least $1 - \delta$.
Note that setting $K>1$ can only improve the estimate.

\subsection{Theoretical bound on the variance of the purity estimate}\label{subsec:variance-bound-depends-on-purity-CS-protocol}
To derive an upper bound on the variance of the purity estimate given by the classical shadows protocol, and show its dependency on the purity of the target state, we exploit results from the Supplementary Material of the original paper on the classical shadows protocol \cite{huang_predicting_2020}.

First, recall that the purity estimate is given by

\begin{equation}
    \hat{P}_{\text{Shadows}}^{(M,K)} \coloneqq \frac{2}{M(M-1)} \sum_{m < m'} Tr[\hat{\rho}^{(m)}\hat{\rho}^{(m')}] = \binom{M}{2}^{-1} \sum_{m < m'} Tr[\hat{\rho}^{(m)}\hat{\rho}^{(m')}] \,,
\end{equation}
where
\begin{equation}
    \hat{\rho}^{(m)} = \frac{1}{K}\sum_{k=1}^K \hat{\rho}^{(m,k)} = \frac{1}{K}\sum_{k=1}^K \bigotimes_{n=1}^{N} \left( 3 \left(U_n^{(m)} \right)^\dagger \ket{s_n^{(m,k)}}\bra{s_n^{(m,k)}} U_n^{(m)} - I \right)\,.
\end{equation}

Performance bounds on the variance derived in Ref. \cite{huang_predicting_2020} are for the single-shot limit $K=1$. The variance of our purity estimate in the $K>1$ case may be upper bounded by the single-shot variance as follows:
\begin{align}
    V[\hat{P}_{\text{Shadows}}^{(M,K)}] &= V\left[\frac{1}{K^2} \sum_{k,k'} \hat{P}(k,k')\right] \quad \text{where} \quad \hat{P}(k,k') \equiv \frac{2}{M(M-1)} \sum_{m<m'} \Tr[\hat{\rho}^{(m,k)}\hat{\rho}^{(m',k')}]\\
    &= \frac{1}{K^4} V\left[\sum_{k,k'} \hat{P}(k,k')\right] \\
    &\leq \frac{1}{K^4} K \sum_{k} V\left[\sum_{k'} \hat{P}(k,k')\right]\\
    &\leq \frac{1}{K^4} K^2 \sum_{k,k'} V\left[ \hat{P}(k,k')\right]\\
    &=\frac{1}{K^2} \sum_{k,k'} V\left[ \hat{P}(1,1)\right]\\
    &=  V\left[ \hat{P}_{\text{Shadows}}^{(M,1)} \right]\,,
\end{align}
where we have used twice that the variance of a sum of random variables is upper bounded by the number of elements in the sum times the sum of the variances of those random variables. Indeed, for $n$ discrete random variables $X_k$, we have

\begin{align}
    V\left[\sum_{k=1}^n X_k\right] &= \sum_{i,j} Cov\left[X_i, X_j\right]\\
    &\leq \sum_{i,j} \sqrt{V[X_i]V[X_j]}\\
    &\leq \sum_{i,j} \frac{V[X_i] + V[X_j]}{2}\\
    &=\frac{1}{2}  (n\sum_iV[X_i] + n\sum_j V[X_j])\\
    &=n \sum_k V[X_k]
\end{align}
where we have used Cauchy-Schwarz inequality to upper bound the covariance with variances, as well as the fact that $\sqrt{xy}\leq (x+y)/2$ for positive $x,y$.

Now using a similar derivation as in Eq. (S65) of the Supplementary Material of Ref. \cite{huang_predicting_2020}, the single-shot variance satisfies
\begin{align}
    V[\hat{P}_{\text{Shadows}}^{(M,1)}] &= \binom{M}{2}^{-2} \sum_{m < m'} (\mathbf{E}\left[  Tr[\hat{\rho}^{(m)}\hat{\rho}^{(m')}] ^2 \right] - {\Tr[\rho^2]}^2 ) \\ 
    &+ 2 \binom{M}{2}^{-2} \sum_{m < m'} \sum_{n \neq m, m'} (\mathbf{E}\left[  Tr[\hat{\rho}^{(m)}\hat{\rho}^{(m')}]  Tr[\hat{\rho}^{(m)}\hat{\rho}^{(n)}] \right] - {\Tr[\rho^2]}^2 )\\
    &= \binom{M}{2}^{-2} \sum_{m < m'} V\left[\Tr[\hat{\rho}^{(1)}\hat{\rho}^{(2)}]\right] + 2 \binom{M}{2}^{-2} \sum_{m < m'} \sum_{n \neq m, m'} V\left[\Tr[\hat{\rho}^{(1)}\rho]\right]\\
    &= \binom{M}{2}^{-1}  V\left[\Tr[\hat{\rho}^{(1)}\hat{\rho}^{(2)}]\right] + 2 \binom{M}{2}^{-1} (M-2)  V\left[\Tr[\hat{\rho}^{(1)}\rho]\right]\\
    &\leq \frac{4}{M^2} V\left[\Tr[\hat{\rho}^{(1)}\hat{\rho}^{(2)}]\right] + \frac{4}{M}  V\left[\Tr[\hat{\rho}^{(1)}\rho]\right] \\ 
    &= \frac{4}{M^2} V\left[\Tr[\hat{\rho}^{(1,1)}\hat{\rho}^{(2,1)}]\right] + \frac{4}{M}  V\left[\Tr[\hat{\rho}^{(1,1)}\rho]\right]\,,
\end{align}
where in the line before the last one we have used the fact that $\binom{M}{2}^{-1} \leq \frac{4}{M^2}$ and $2 \binom{M}{2}^{-1} (M-2) \leq \frac{4}{M}$ for $M\geq 2$.

Noticing that 
\begin{equation}
    V\left[\Tr[\hat{\rho}^{(1,1)}\hat{\rho}^{(2,1)}]\right] =  V\left[\Tr[ O_2 \hat{\rho}_2]\right] \quad \text{where} \quad O_2=S \quad \text{and} \quad \hat{\rho}_2 = \hat{\rho}^{(1,1)}\otimes\hat{\rho}^{(2,1)}\,,
\end{equation}
and
\begin{equation}
    V\left[ \Tr[\hat{\rho}^{(1,1)}\rho]\right] = V\left[\Tr[ O_1 \hat{\rho}_1]\right] \quad \text{where} \quad O_1=\rho \quad\text{and} \quad \hat{\rho}_1 = \hat{\rho}^{(1,1)}\,,
\end{equation}
we can use Lemma (S1) of the Supplementary Material of Ref. \cite{huang_predicting_2020}, to upper bound those variance terms with the shadow norm squared:
\begin{equation}
    V\left[\Tr[\hat{\rho}^{(1,1)}\hat{\rho}^{(2,1)}]\right] \leq ||O_{2,0}||_{\text{shadow}}^2 \quad \text{where} \quad O_{2,0}=S - \Tr[S]\frac{I_{2n}}{2^{2n}}=S - \frac{I_{2n}}{2^{n}}\,,
\end{equation}
and
\begin{equation}
    V\left[ \Tr[\hat{\rho}^{(1,1)}\rho]\right] \leq ||O_{1,0}||_{\text{shadow}}^2 \quad \text{where} \quad O_{1,0} = \rho - \Tr[\rho]\frac{I_n}{2^n} = \rho - \frac{I_n}{2^n}\,.
\end{equation}
Finally, noticing that the SWAP operator has locality $2n$, and for a state $\rho$ of unknown locality upper bounded by $n$, we use the last line of the proof of Proposition (S3) from the Supplementary Material of Ref. \cite{huang_predicting_2020}, to upper bound each shadow norm as a function of the two-norm (Hilbert-Schmidt norm) squared ($||O||_2^2 = \Tr[O^2]$):
\begin{equation}
V\left[\Tr[\hat{\rho}^{(1,1)}\hat{\rho}^{(2,1)}]\right] \leq 2^{2n} ||O_{2,0}||_2^2 = 4^n \Tr[O_{2,0}^2] = 4^n \Tr[S^2 - 2S\frac{I_{2n}}{2^{n}} + \frac{I_{2n}}{2^{2n}}] = 4^n (4^{n} - 1) \leq 16^n \,,
\end{equation}
and
\begin{equation}
    V\left[ \Tr[\hat{\rho}^{(1,1)}\rho]\right] \leq 2^n||O_{1,0}||_2^2 = 2^n\Tr[O_{1,0}^2] = 2^n\Tr[\rho^2 - 2\frac{\rho}{2^n} + \frac{I_n}{2^{2n}}] =  2^n\left(\Tr[\rho^2] - \frac{1}{2^n}\right) \leq 2^n \Tr[\rho^2]\,,
\end{equation}
where it may be interesting to note that the last term of the above equation would have become $2^{2k-n}\Tr[\rho^2]$ if we had decided to take into account the locality $k$ of the state $\rho$ in the derivation.

Combining everything together, we conclude that
\begin{equation}
    V[\hat{P}_{\text{Shadows}}^{(M,K)}] \leq V[\hat{P}_{\text{Shadows}}^{(M,1)}]
    \leq \frac{4}{M^2} 16^n + \frac{4}{M}  2^n \Tr[\rho^2]
\end{equation}

Therefore, the smaller the purity of the target state (i.e., the more mixed the state), the smaller the upper bound on the variance of the purity estimate outputted by the classical shadows protocol. Choosing $M$ of order $4^n$, we see that the effect of the purity of the state in the upper bound on the variance should become less noticeable as we consider larger system sizes and the $16^n$ term becomes dominant.

\subsection{Systematic error in the derandomized version of the classical shadows protocol}\label{subsec:systematic_error_derand_CS}

\begin{figure*}
     \centering
     \begin{subfigure}[b]{\textwidth}
         \centering
         \includegraphics[width=\textwidth]{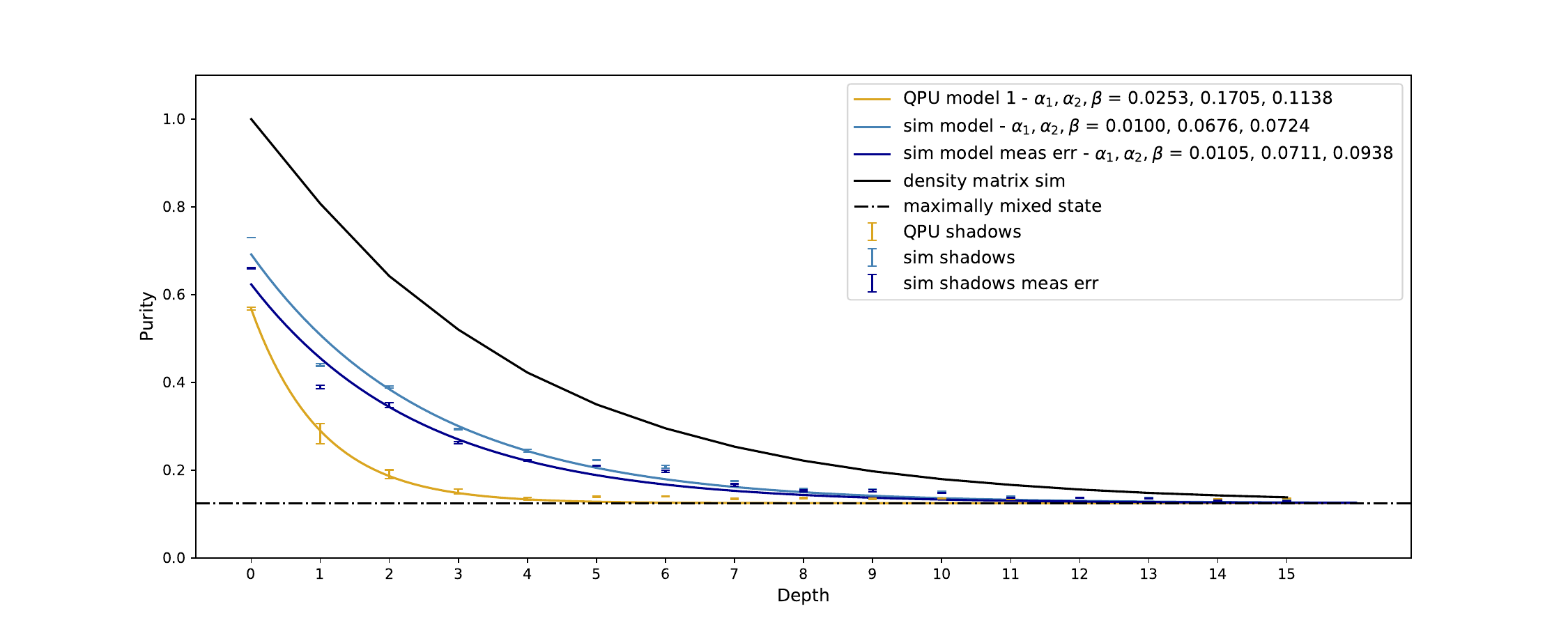}
         \caption{Purity}
         \label{subfig:QPU_run_n=3_classim_and_QPU_CS-R_density-purity}
     \end{subfigure}
     \begin{subfigure}[b]{\textwidth}
         \centering
         \includegraphics[width=\textwidth]{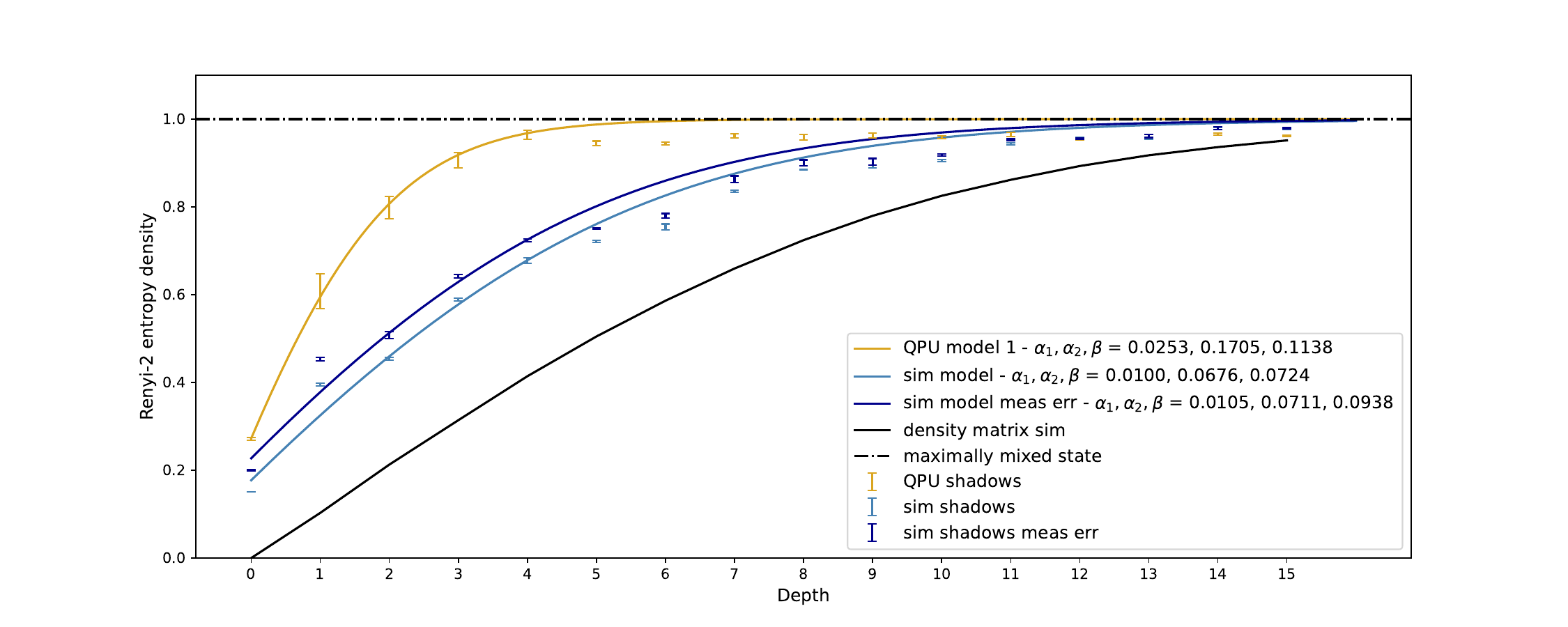}
         \caption{Renyi entropy density}
         \label{subfig:QPU_run_n=3_classim_and_QPU_CS-R_density}
     \end{subfigure}
        \caption{\textbf{QPU results and validation of our heuristic model.} We consider a VQA circuit as in Fig.~\ref{fig:ideal_hardware_efficient_param_circuit} applied to $n=3$ qubits, with fixed circuit parameters (fixed random seed \texttt{np.random.seed(837)}). Gold error bars give the purity and Renyi-2 entropy evolution as functions of circuit depth on Rigetti's QPU Aspen-M-3 using the derandomized version of the protocol (with $M=3^3=27$ measurement settings considered in order). Each error bar was obtained by running the classical shadows protocol three times, and computing the average estimate and the standard deviation over those three samples. Results obtained in the same setting but using a local depolarizing noise model ($p_1=0.008$ and $p_2=0.054$ from calibration data of Aspen-M-3) correspond to the blue error bars; dark (light) blue error bars were obtained with (without) readout errors from calibration data. The black solid lines show the purity and Renyi-2 entropy density evolution obtained via density matrix simulation under our local depolarizing noise model. The black horizontal dash-dotted lines correspond to the purity or entropy density of the maximally mixed state $\sigma_0\coloneqq I/2^n$. Fit of our global depolarizing heuristic model, where we have imposed that $\alpha_2/\alpha_1=p_2/p_1$, corresponds to the solid gold and blue lines.}
        \label{fig:QPU_run_n=3_classim_and_QPU_CS-systematic-error}
\end{figure*}

Changing circuits is relatively costly in execution time for some platforms like superconducting qubits. Since our entropy accumulation QPU data were limited in number of qubits ($n=3$) and therefore we were expecting to see the occurrence of most measurements and in similar number, we initially opted for a derandomization of the measurement choice where we consider one by one all of the $3^3=27$ possible measurement settings with $K=1000$ classical snapshots of the output state for each of them. The equivalent of Fig.~\ref{fig:QPU_run_n=3_classim_and_QPU_CS} for QPU entropy accumulation but in the case of derandomized measurements of the classical shadows protocol corresponds to Fig.~\ref{fig:QPU_run_n=3_classim_and_QPU_CS-systematic-error}, where we can observe the presence of a systematic bias in the purity or Renyi-2 entropy density estimate provided by the classical shadows protocol.

\subsection{Systematic error for low number of samples}\label{subsec:systematic_error_low_M_CS}

\begin{figure}
    \centering
    \includegraphics[width=0.7\linewidth]{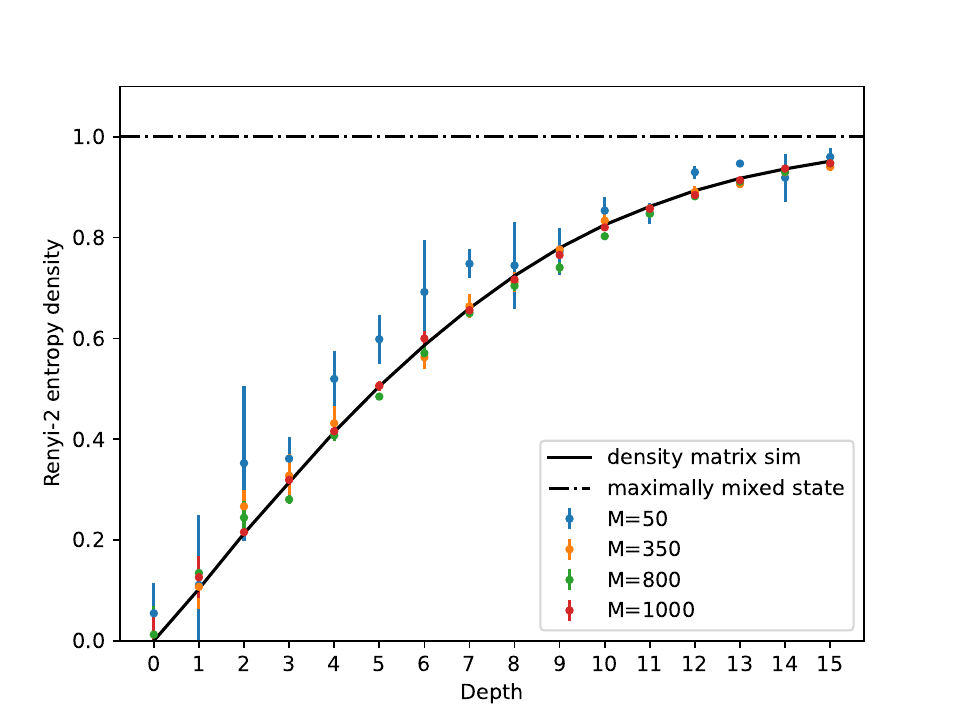}
    \caption{\textbf{Systematic error in the classical shadows estimate for low sample complexity.} We consider an ($n=3$)-qubit VQA circuit as in Fig.~\ref{fig:ideal_hardware_efficient_param_circuit}, with fixed circuit parameters, under local depolarizing noise (with parameters $p_1=0.008$ and $p_2=0.054$ from calibration data of Aspen-M-3). The solid black line corresponds to the Renyi-2 entropy density evolution of the circuit output for different circuit depths. We also estimate the entropy density using the Pauli version of the classical shadows protocol for a varying number of measurement settings $M\in \{ 50; 350; 800; 1000\}$ but with fixed parameters $K=1000$ and $N_g=5$ and each time running the protocol three times to obtain error bars. The black horizontal dash-dotted line corresponds to the entropy density of the maximally mixed state $\sigma_0\coloneqq I/2^n$.}
    \label{fig:CS_sys_err_low_M}
\end{figure}

When using the classical shadows protocol for Renyi-2 entropy or purity estimation, it is crucial to choose a sufficiently large number of samples, $MK$, and more precisely number of measurement settings, $M$, to avoid any systematic error in the estimate. We illustrate the importance of the choice of value for $M$ in Fig.~\ref{fig:CS_sys_err_low_M}. In that figure, as an example, we consider an ($n=3$)-qubit VQA circuit as in Fig.~\ref{fig:ideal_hardware_efficient_param_circuit}, with fixed circuit parameters, under local depolarizing noise from calibration data of Aspen-M-3. Then, we estimate the output Renyi-2 entropy density as a function of circuit depth using the classical shadows protocol with all parameters fixed except the number of measurement settings $M$. We plot the resulting entropy density estimate for $M\in \{ 50; 350; 800; 1000\}$ where each value corresponds to a different colour of error bars in the figure. As a reference, we show the exact entropy density evolution as a solid black line. We observe that, for very small $M$ value (e.g., $M=50$), the estimate is above the exact value. As $M$ increases, the systematic error decreases, and the estimate converges to the exact value (e.g., $M=350$). A minor systematic error can sometimes be observed for larger $M$ values (e.g., $M=800$) where the estimated entropy evolution is slightly below the exact curve; however, we have observed that increasing $M$ even further (e.g., $M=1,000$) removes this small deviation.

%% file: app-QAdv_benchmarking.tex
\section{Circuit size boundary for reachable quantum advantage}\label{app:circuit_size_boundary_condition_for_qadv}

In this section, we prove that our bound on the circuit size boundary for reachable quantum advantage (Eq.~\ref{eq:circuit-size-qadv-condition-large-n}) is tighter than that of Ref. \cite{stilck_franca_limitations_2021} for a large number of qubits.

The circuit-size condition for classical superiority from Ref. \cite{stilck_franca_limitations_2021} is 
\begin{equation}
    D \geq \frac{1}{2\alpha}\ln{(1-c)^{-1}}\,,
\end{equation}
while at $n \rightarrow \infty$, ours is
\begin{equation}
    D \geq \frac{1}{2\alpha}c \ln{(2)} \,.\label{eq:circuit-size-qadv-condition-large-n}
\end{equation}

Let $f(c) \coloneqq c \ln{(2)} - \ln{((1-c)^{-1})} = c \ln{(2)} + \ln{(1-c)}, \forall c \in [0,1]$.
We have $\frac{df}{dc}(c) = \ln{(2)} - \frac{1}{1-c}$, which can be shown to satisfy $\frac{df}{dc}(c) < 0, \forall c \in [0,1]$. Therefore, $f$ is a strictly decreasing function on its domain $c \in [0,1]$. Since $f(0) = 0$, we conclude that $f(c) \leq 0$ for all values of $c \in [0,1]$.
This means that for large problem sizes, i.e., $n \rightarrow \infty$, quantum advantage will be lost at a shorter depth (Eq.~\eqref{eq:circuit-size-qadv-condition-large-n}) than that predicted in Ref. \cite{stilck_franca_limitations_2021}. Note that the above argument remains valid even for the more realistic case where the entropy density threshold $c$ keeps a dependency with the system size $n$, i.e., $c=c(n)$. This is because the value $c(n)$ remains the same input for both bounds and $f(c) \leq 0$ for all values of $c \in [0,1]$ remains true irrespective of the variables that may define $c$ ($n$ in our case or others).

%% file: app-Aspen-M-3-calibration-data.tex
\section{Rigetti's Aspen-M-3 and link between calibration data and noise parameters}\label{app:calibration_data_Aspen-M-3}

We link noise parameters of the noise model used for numerical simulations with typical calibration data given for a quantum device.

\begin{figure}[h!]
    \centering
    \includegraphics[width=.2\textwidth]{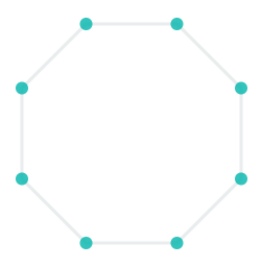}
    \caption{Eight-qubit Aspen-M-3 QPU}
    \label{fig:Aspen-M-3}
\end{figure}

All QPU results shown in this article were obtained on Rigetti's eight-qubit Aspen-M-3 device (see Fig.~\ref{fig:Aspen-M-3}). For one-qubit circuits, qubit $\#100$ was used; for two-qubit circuits, qubits $\#100, 101$ were used; for three-qubit circuits, $\#100, 101, 102$; and so on. Calibration data from the day of our circuit runs on the QPU are provided in Table \ref{tab:Aspen-M-3_calibration_data}. Typical gate times for the device are $T_{gate,1}=40ns$ and $T_{gate,2}=240ns$ for single-qubit and two-qubit gates. The measurement process can take more than $T_{meas}=1\mu s$ \cite{pyquilreadoutnoise}.

\begin{table}[h!]
\begin{tabular}{lllllllll}
\hline
Pair     & fXY      & fXY std err & fCZ    & fCZ std err   & Avg T1 (µs) & Avg T2 (µs)        & Avg fActiveReset & Avg fRO \\\hline
100--101 & 0.9614   & 0.007829    & 0.9598 & 0.0058        & 19.3         & 33.22               & 0.9943           & 0.941   \\
100--107 & 0.8553   & 0.007008    & 0.9769 & 0.005897      & 27.2         & 45.47               & 0.996            & 0.944   \\
101--102 & 0.9626   & 0.009064    &        &               & 21.36        & 32.92               & 0.996            & 0.9165  \\
102--103 & 0.8409   & 0.01479     & 0.9711 & 0.007058      & 26.6         & 29.48               & 0.9858           & 0.9135  \\
103--104 & 0.8534   & 0.006161    & 0.909  & 0.003757      & 31.31        & 43.49               & 0.8827           & 0.9455  \\
104--105 & 0.812    & 0.007665    & 0.8917 & 0.004528      & 29.04        & 35.61               & 0.8945           & 0.945   \\
105--106 & 0.9756   & 0.003552    &        &               & 27.46        & 15.55               & 0.9885           & 0.9735  \\
106--107 & 0.9828   & 0.002881    & 0.9948 & 0.002183      & 32.39        & 39.13               & 0.9888           & 0.98    \\\hline
         &          &             &        &               &              &                     &                  &         \\\hline
Qubit    & T1 (µs) & T2 (µs)    & f1QRB  & f1QRB std err & f1Q sim. RB  & f1Q sim. RB std err & fActiveReset     & fRO     \\\hline
100      & 11.82    & 20.1        & 0.997  & 0.000175      & 0.9957       & 0.000109            & 0.9955           & 0.908   \\
101      & 26.78    & 46.34       & 0.9997 & 0.000382      & 0.9989       & 5.61E-05            & 0.993            & 0.974   \\
102      & 15.95    & 19.5        & 0.9977 & 0.00014       & 0.9959       & 0.000148            & 0.999            & 0.859   \\
103      & 37.25    & 39.46       & 0.9953 & 0.000827      & 0.9869       & 0.000571            & 0.9725           & 0.968   \\
104      & 25.36    & 47.52       & 0.8209 & 0.04649       & 0.8665       & 0.0576              & 0.793            & 0.923   \\
105      & 32.72    & 23.71       & 0.9984 & 6.81E-05      & 0.9949       & 0.000161            & 0.996            & 0.967   \\
106      & 22.2     & 7.401       & 0.999  & 4.9E-05       & 0.9973       & 0.000185            & 0.981            & 0.98    \\
107      & 42.58    & 70.85       & 0.9995 & 6.44E-05      & 0.9989       & 2.96E-05            & 0.9965           & 0.98   \\\hline
\end{tabular}
\caption{Aspen-M-3 calibration data on 9 December 2023 at 9:59pm (UTC+00:00)}
\label{tab:Aspen-M-3_calibration_data}
\end{table}

\subsection{Gate noise}

\subsubsection{Local depolarizing noise}

Assuming local depolarizing noise as the noise model on the device for classical simulations, it is possible to link one-qubit and two-qubit gate fidelities ($F_1$ and $F_2$) from calibration data to one-qubit and two-qubit depolarizing probabilities ($p_1$ and $p_2$) as follows:

\begin{align}
    p_1 &= 2 (1-F_1) \,,\\
    p_2 &= \frac{4}{3} (1-F_2) \,,
\end{align}

where we have used the fact that the fidelity between a pure state $\ket{\Psi}$ and a mixed state $\rho$ is $F=\bra{\Psi} \rho \ket{\Psi}$ and taking for $\ket{\Psi}$ a pure state evolved with some ideal gate $U$, and for $\rho$ a pure state evolved with $U$ followed by a depolarizing channel (e.g., see Ref. \cite{gaebler2012randomized}). For our classical simulations, we considered the median sim RB fidelity (one-qubit gate fidelity) and the median CZ fidelity (two-qubit gate fidelity) in agreement with the two-qubit gates of our VQA ansatz.

\subsubsection{Amplitude damping ($T_1$ relaxation) and local depolarizing noise}
In the case where gate noise in the device can be modeled via a \textit{combination} of local depolarizing noise followed by amplitude damping, one may link one-qubit and two-qubit gate fidelities ($F_1$ and $F_2$), the relaxation time $T_1$ and single-qubit and two-qubit gate times $T_{gate, 1}$ and $T_{gate, 2}$---as obtained from calibration data---to the corresponding one-qubit and two-qubit depolarizing probabilities ($p_1$ and $p_2$) as well as the amplitude damping probabilities ($\gamma_1$ and $\gamma_2$), respectively, as follows:
\begin{align}
    \gamma_1 &= 1 - e^{-\frac{T_{gate,1}}{T_1}}\\
    \gamma_2 &= 1 - e^{-\frac{T_{gate,2}}{T_1}}\\
    p_1 &= 1 - \frac{2(3 F_1 - 1) - x^2}{1 + 2x}\\
    p_2 &= 1 - \frac{4(5 F_2 - 1) - 1}{y (4 + 6y + 4y^2 + y^3)}\,,
\end{align}
where $x = \sqrt{1 - \gamma_1}$ and $y = \sqrt{1 - \gamma_2}$. Here, we have used Eq. (3) from Ref. \cite{nielsen2002simple}---which links average gate fidelity and entanglement fidelity---to easily express average gate fidelities $F_1$ and $F_2$ as functions of noise parameters $p_1, \gamma_1$ and $p_2, \gamma_2$, respectively. In particular, we defined single-qubit and two-qubit noise channels:
\begin{align}
    \mathcal{E}_1 &= A_{\gamma_1} \circ D_{p_1}\\
    \mathcal{E}_2 &= A_{\gamma_2} \circ D_{p_2}\,,
\end{align}
where the single-qubit depolarizing and amplitude damping channels are
\begin{align}
    D_{p_1}: \rho &\mapsto (1-p_1) \rho + p_1 \frac{I}{2} \\
    A_{\gamma_1}: \rho &\mapsto E_0 \rho E_0^\dagger + E_1 \rho E_1^\dagger
\end{align}
and the two-qubit depolarizing and amplitude damping channels are
\begin{align}
     D_{p_2}: \rho &\mapsto (1-p_2) \rho + p_2 \frac{I}{4} \\
    A_{\gamma_2}: \rho &\mapsto F_0\otimes F_0 \rho (F_0\otimes F_0)^\dagger + F_0\otimes F_1 \rho (F_0\otimes F_1)^\dagger + F_1\otimes F_0 \rho (F_1\otimes F_0)^\dagger + F_1\otimes F_1 \rho (F_1\otimes F_1)^\dagger
\end{align}
with Kraus operators
\begin{equation}
    E_0 = \begin{pmatrix}
        1 & 0\\
        0 & \sqrt{1-\gamma_1}
        \end{pmatrix}, 
        \quad 
    E_1 = \begin{pmatrix}
        0 & \sqrt{\gamma_1}\\
        0 & 0
        \end{pmatrix},
        \quad
    F_0 = \begin{pmatrix}
        1 & 0\\
        0 & \sqrt{1-\gamma_2}
        \end{pmatrix},
        \quad
    F_1 = \begin{pmatrix}
        0 & \sqrt{\gamma_2}\\
        0 & 0
        \end{pmatrix}\,.
\end{equation}

\subsection{Readout noise}
\subsubsection{Amplitude damping ($T_1$ relaxation)}

In classical simulations with assumed $T_1$ relaxation in the measurement stage, we add a layer of single-qubit amplitude damping channels with error probability $\gamma_{meas}=1 - e^{-\frac{T_{meas}}{T_1}}=0.05$ (based on calibration data of Aspen-M-3) just before the computational basis measurement. More precisely, the single-qubit amplitude damping channel in the measurement stage is defined as
\begin{align}
    A_{\gamma_{meas}}: \rho &\mapsto G_0 \rho G_0^\dagger + G_1 \rho G_1^\dagger\,,
\end{align}
where
\begin{equation}
    G_0 = \begin{pmatrix}
        1 & 0\\
        0 & \sqrt{1-\gamma_{meas}}
        \end{pmatrix}, 
        \quad 
    G_1 = \begin{pmatrix}
        0 & \sqrt{\gamma_{meas}}\\
        0 & 0
        \end{pmatrix}.
\end{equation}

\subsubsection{Classical bit flip}

For classical simulations where readout errors were included, we estimated the probabilities $P(0|1)$ and $P(1|0)$ on Aspen-M-3, and obtained
\begin{equation}
    P(0|1) = 0.03484722222 \quad P(1|0) = 0.022177777\,,
\end{equation}
where $P(m|n)$ is the probability of obtaining outcome $m$ instead of $n$ due to readout error. In our heuristic model from Eq.~\eqref{eq:purity_model_CS_measerr}, adapted to take into account readout errors, $p_m$ is related to $P(0|1)$ and $P(1|0)$ in the case where those have similar values (symmetric readout error).
For the more accurate classical simulations with added $T_1$ relaxation effects, we take $P(0|1) = P(1|0) = 0.03$ which is a reasonable assumption based on the observed asymmetry and to make the classical simulation more practical since standard publicly available calibration data usually provides readout fidelities $F_{RO} = 1-P(0|1) = 1-P(1|0)$, which do not give information on the error asymmetry.